\documentclass[aps, prd, twocolumn, a4paper]{revtex4-1}
\usepackage{ulem}
\usepackage{color}
\usepackage{appendix}
\usepackage{graphicx}
\usepackage{epsfig}
\newcommand{\be}{\begin{equation}}
\newcommand{\ee}{\end{equation}}
\newcommand{\ba}{\begin{eqnarray}}
\newcommand{\ea}{\end{eqnarray}}
\newcommand{\nn}{\nonumber\\}
\begin{document}
\title{Thermal Relaxation, Electrical Conductivity and Charge Diffusion in  a Hot QCD Medium}
\author{Sukanya Mitra}
\email {sukanyam@iitgn.ac.in} 
\author{Vinod Chandra}
\email {vchandra@iitgn.ac.in}
\affiliation{Indian Institute of Technology Gandhinagar,  Gandhinagar-382355, Gujarat, India}
\begin{abstract}
The response of electromagnetic (EM) fields that are produced in non-central heavy-ion collisions 
to electromagnetically charged quark gluon plasma can be understood in terms of charge transport and 
charge diffusion in the hot QCD medium. This article presents a perspective on these processes 
by investigating the temperature behavior of the related transport coefficients, {\it viz.} electrical conductivity and 
the charge diffusion coefficients along with charge susceptibility. In the process of estimating them, 
thermal relaxation times for quarks and gluons have been determined first.
These transport coefficients have been studied by solving the relativistic transport equation in the Chapman-Enskog method. For the analysis, $2\rightarrow 2$,
quark-quark, quark-gluon and gluon-gluon scattering processes are taken into account along with an effective description of hot 
QCD Equations of state (EOSs) in terms of temperature dependent effective fugacities of quasi-quarks (anti-quarks) and quasi-gluons. 
Both improved perturbative hot QCD EOSs  at high temperature and a lattice QCD EOS are included for the analysis. The hot QCD 
medium effects entering through the quasi-particle momentum distributions along with an effective coupling, are seen to have 
significant impact on the temperature behavior of these transport parameters along with the thermal relaxation times for the 
quasi-gluons and quasi-quarks.\\

\\

\end{abstract}
\maketitle
\section{Introduction}
Quantum chromodynamics(QCD)--the underlying theory of strong interaction in nature, predicts a deconfined state of the nuclear matter at high temperature (higher than QCD transition temperature  $T_c$). 
Relativistic heavy-ion collision experiments at RHIC, BNL and  LHC, CERN have reported the presence of near perfect liquid like hot nuclear matter~\cite{expt_rhic, expt_lhc} which turns out to be  strongly coupled quark-gluon plasma (QGP).
The QGP possess a very tiny value of the shear viscosity to entropy density ratio, $\eta/S$ (a few times the KSS bound~\cite{kss}).  
The $\eta/S$ has a lower bound near the transition temperature $T_c$ as shown by several studies on QCD matter based on various 
approaches and the shear viscosity to entropy density ratio for the QGP is found to be lowest among all the known fluids~\cite{lowvisc, lacey}.
In contrast,  the  bulk viscosity to entropy density ratio, $\zeta/S$ shows an upper bound with large values~\cite{bulk}.

There has been growing interest in understanding  the impact of strong electro-magnetic (EM) fields  that are produced during the initial stages of the 
non-central heavy ion collisions~\cite{em_rhic}, while investigating the hadronic observables at the later stages of the collisions.
The impact of the EM field will be certainly dependent on the strength of the fields at the later stages as they seen to decay quite 
rapidly~\cite{em_decay}. The response of such EM fields (electrical) to the 
electromagnetically charged QGP  can be understood in terms of the electrical conductivity, $\sigma_{el}$ which characterizes the transport of the $U(1)$ conserved charge in the presence of the 
gradient of a charge chemical potential. 
There are several recent attempts to understand the $\sigma_{el}$ in the context of EM field in RHIC~\cite{em_decay, lerry, kharz, satw}.
The electrical conductivity in the context of charge fluctuations in heavy-ion collisions is investigated in~\cite{charge_fluc}. 
Moreover, there are recent proposals to extract the electrical conductivity from the flow parameters in heavy ion collisions~\cite{flow}.

The other relevant, associated  physical process is the charge diffusion, that is being quantified by the charge diffusion coefficient, $D$. 
The electrical conductivity, $\sigma_{el}$ and the diffusion coefficient $D$
are related by the famous Einstein relation through the charge susceptibility, $\chi$.

The prime focus of this work is to estimate all these quantities for a hot QCD/QGP medium (their temperature dependence) which is characterized by an
effective quasi-particle model.  At this juncture, one needs to demarcate between the comoving frames while modeling expanding QGP medium, 
one with moving charge density and other one with energy density.  This requires the 
introduction of thermal conductivity which characterizes the flow associated with the transport of energy in response to the temperature 
gradient relative to locally comoving frames with the charge density. There are some recent attempts in the 
direction ~\cite{Greif_heat,Dobado}, however, our work in this manuscript  does not involve any such investigation.  As there are two equivalent approaches to estimate the transport coefficients of the hot 
QCD/QGP medium, {\it viz.}, the linear response theory where one could relate them to the  
current-current spectral functions in thermal equilibrium through the Green-Kubo formulae~\cite{green-kubo}, the other one is to solve a 
linearized transport equation in the presence of electric field along with an appropriate collision term (again linearized) and invoke 
pertinent equations of motion (for example Maxwell's equations in the case of  electrical conductivity and electrical permittivities). 
The former approach best suited to lattice QCD estimations of the electrical conductivity and charge diffusion coefficient. There are 
several attempts from lattice QCD side to obtain the temperature dependence of the conductivity and charge diffusion 
coefficient~\cite{Arts-2007,Amato,Arts-2015,Gupta,Brandt,Francis, Buividovich} along with charge susceptibility~\cite{Giudice}. 
The present work follows the latter approach based on the linearized  transport equation. 

There are a number of estimations of electrical conductivity ($\sigma_{el}$) by different approaches available in current literatures.
In Ref. ~\cite{Xu-Greiner,Greco-1,Greco-2} the electrical conductivity has been estimated by solving the relativistic transport equation.
In Ref. ~\cite{Cassing,Cassing1} the $\sigma_{el}$ has been studied using off-shell parton-hadron string dynamics transport approach
for an interacting system. In  Ref.~\cite{Qin} $\sigma_{el}$ as a function of temperature has been estimated using
the maximum entropy method (MEM). Electrical conductivity along with diffusion coefficient and charge susceptibility has been
estimated employing holographic technique in Ref. ~\cite{Finazzo}. Recently the electrical conductivity has been studied
including the momentum space anisotropy also  ~\cite{Binoy}. In the hadronic sector also 
these quantities have been investigated lately. In Ref. \cite{Fraile} the electrical conductivity has been evaluated for a pion gas
where in Ref.~\cite{Denicol} the same has been studied for hot hadron gas.

While setting up an appropriate transport equation with a collision term for the determination of $\sigma_{el}$ and $D$ for the QGP medium, one must have a reliable modeling of the equilibrium state of the medium.
 To that end, quasi-particle descriptions of hot QCD medium play an important role. We employ such a model which is based on mapping the hot QCD medium effects  encoded in the equation of state to the non-interacting/weakly interacting 
 quasi-parton degrees of freedom with temperature dependent effective fugacity parameter~\cite{chandra_quasi1,chandra_quasi2}. Further, the model can be understood in terms of  renormalization of charges of quasi-partons in the 
 hot QCD medium. This enables us to define an effective coupling constant in hot QCD medium. The effective coupling thus obtained is employed in our analysis.
 
 The manuscript is organized as follows. Section II, offers mathematical formalism to  determine thermal relaxation times 
 for quasi-partons (quasi-quarks/antiquarks and quasi-gluons) followed by the analytical estimations of the transport coefficients,
 $\sigma_{el}$, $D$ and $\chi$. Section III, deals with important predictions of the transport coefficients mentioned and 
 related discussions. Finally, in Section IV, the conclusions and outlook of the work are presented.

\section{Formalism}
\subsection{Quasi-particle description of hot QCD medium}
Realization of hot QCD medium effects in terms of effective quasi-particle models has been there since the
last few decades. In fact, there are various quasi-particle descriptions {\it viz.}, effective mass  models~\cite{effmass1, effmass2}, effective mass models with Polyakov loop~\cite{polya}, 
NJL (Nambu Jona Lasinio) and PNJL (Polyakov loop extended Nambu Jona Lasinio) based effective models~\cite{pnjl}, and effective fugacity quasi-particle description of hot QCD (EQPM)~\cite{chandra_quasi1, chandra_quasi2}. The present analysis considers  the 
EQPM (Effective quasiparticle model)for the investigations on the properties of hot and dense medium  in RHIC.  

 There are a number of estimations for different transport coefficients available in current literature which employ various quasiparticle models~\cite{Bluhm,chandra_eta, chandra_etazeta}.
In  Ref. ~\cite{Bluhm},   $\eta$ and $\zeta$ have been evaluated  for pure gluon plasma employing the effective mass quasi-particle model. On the other hand, in Refs. ~\cite{chandra_eta, chandra_etazeta}, $\eta$ and $\zeta$ are obtained in 
gluonic as well as matter sector.  Refs.~ \cite{PJI,Mkap},  presented the quasi-particle theory 
of $\eta$ and $\zeta$ and their estimations for the hadronic sector.  The thermal conductivity
has also been studied, in addition to the viscosity parameters~\cite{Mkap},  within the effective mass model. 
In  Ref.~\cite{Greco}, the ratio of electrical conductivity to shear viscosity has been investigated
within the framework of quasiparticle approach as well. However, 
these model calculations are not able to exactly reproduce the shear and bulk
viscosities phenomenologically extracted from the
hydrodynamic simulations of the QGP \cite{Ryu, Denicol1}, consistently agreeing with different
experimental observables measured like the multiplicity, transverse momentum spectra
 and the integrated flow harmonics of charged hadrons. Nevertheless, 
 these quasiparticle approached could be useful in the equilibrium modeling of the 
 hot QCD/QGP. The  predictions based on these model  are still  useful in the sense 
of estimating some possible values these transport coefficients 
from some theoretical models that can considerably describe
the interacting system created in heavy ion collisions.

\subsubsection*{The EQPM}
The EQPM employed here, models the 
hot QCD in terms of effective quasi-partons (quasi-gluons, quasi-quarks/antiquarks). The model is based on the idea of mapping the 
hot QCD medium effects present in the equations of state (EOSs) either computed within improved perturbative QCD or lattice QCD simulations, 
into the effective equilibrium distribution functions for the quasi-partons. The  EQPM for the QCD EOS at $O(g^5)$ (EOS1)  and $O(g^6\ln(1/g)+\delta)$  (EOS2)
have been considered here. Additionally,  the EQPM for the 
recent (2+1)-flavor lattice QCD EoS~\cite{cheng} at physical quark masses (LEOS), has been employed for our analysis. 
There are more recent lattice results with the improved 
actions and refined lattices~\cite{leos1_lat}, for which we need to re-look the model 
with specific set of lattice data specially to define the effective gluonic degrees of freedom.
Therefore, we will stick to the set of lattice data utilized in the model described in Ref.~\cite{chandra_quasi2}.

In either of the cases of above mentioned EOSs,  form of the quasi-parton equilibrium distribution functions, 
 $ f_{eq}\equiv \lbrace f_{g}, f_{q} \rbrace$  (describing the strong interaction effects in terms of effective fugacities $z_{g,q}$) can be written as.
\be
\label{eq1}
f_{g/q}= \frac{z_{g/q}\exp[-\beta E_p]}{\bigg(1\mp z_{g/q}\exp[-\beta E_p]\bigg)}
\ee
where $E_p=|\vec{p}|$ for the gluons and $\sqrt{|\vec{p}|^2+m_q^2}$ for the quark degrees of freedom ($m_q$ denotes the mass of the quarks).
and $\beta=T^{-1}$ denotes inverse of the 
temperature. We use the notation $\nu_g=2(N_c^2-1)$ for gluonic degrees of freedom,
$\nu_{q}=4 N_c N_f$  for $SU(N_c)$ with $N_f$ number of flavors. As we are working 
at zero baryon chemical potential, therefore quark and antiquark distribution functions are the same.
Since the model is valid in the deconfined phase of QCD (beyond $T_c$), therefore, the mass of the light quarks can be neglected as compared to 
the temperature.  As QCD is a $SU(3)$ gauge theory so $N_c=3$ for  our analysis. Noteworthily, the EOS1 which is fully perturbative, is  proposed by Arnold and Zhai~\cite{zhai} and Zhai and Kastening~\cite{kastening}. On the other hand, 
EOS2 which is at $O(g^6\ln(1/g)+\delta)$ is  determined by Kajantie {\it et al.}~\cite{kaj}  while
 incorporating contributions from non-perturbative scales such as  $g T$ and $g^2 T$. 
Notably,  these effective fugacities ($z_{g/q}$) are not merely temperature
dependent parameters that encode the hot QCD medium effects;  they lead to 
non-trivial dispersion relation both in the gluonic and quark sectors as,
\be
\label{eq2}
\omega_{g/q}=E_p+T^2\partial_T ln(z_{g/q}),
\label{epp}
\ee
where $\omega_{g,q}$ denote the quasi-gluon and quasi-quark dispersions (single particle energy) respectively.
The second term in the right-hand side of Eq. \ref{eq2}, encodes the effects from collective excitations of the quasi-partons.

The effective fugacities,  $z_g, z_q$ are not related with any conserved number current in 
the hot QCD medium. They have been merely introduced to encode the hot QCD medium effects in the EQPM. The
 physical interpretation of $z_g$ and $z_q$ emerges from the above mentioned non-trivial dispersion relations. The modified part of 
the energy dispersions in Eq. ({\ref{eq2}) leads to the trace anomaly (interaction measure) in hot QCD and takes care of the thermodynamic consistency condition.  It is straightforward to compute, gluon and quark number densities
and all the thermodynamic quantities such as energy density, entropy, enthalpy {\it etc.} by realizing hot QCD medium in terms of an effective Grand canonical system~\cite{chandra_quasi1,chandra_quasi2}.
Furthermore, these effective fugacities lead to a very simple interpretation of hot QCD medium effects in terms of  an effective Virial expansion. Note that $z_{g,q}$ scales with $T/T_c$, where $T_c$ is the QCD transition temperature.

The number densities, $n_g$ (for gluons), $n_q$ (for quarks, antiquarks) are obtained from Eq. (\ref{eq1}) as, 
\ba
\label{eq3}
n_g&=& \frac{\nu_g}{(2\pi)^3} \int d^3 \vec{p} f_g (\vec{p})\nn
&=& \frac{\nu_g T^3}{\pi^2} PolyLog[3, z_g],
\ea
\ba
\label{eq4}
n_q&=& \frac{\nu_q}{(2\pi)^3} \int d^3 \vec{p} f_q (\vec{p})\nn
&=& \frac{-\nu_q T^3}{\pi^2} PolyLog[3, -z_q].
\ea
The number densities approach to their SB (Stefan Boltzmann) limit only asymptotically ({\it i.e} $ z_{g,q}\rightarrow 1$).
On the other hand, the pressure, $P\equiv P_g+P_q$, Energy density, $\epsilon=\epsilon_g+\epsilon_q$ can be obtained from the relation:
\ba
\label{eq5}
P_{g,q}&=&\mp \nu_{g,q}\int \frac{d^3p}{(2\pi)^3} \ln(1\mp z_{g,q} \exp(-\beta E_p))\nn
&=& \pm \frac{\nu_{g,q} T^4}{ \pi^2} PolyLog[4, \pm z_{g,q}],
\ea
\ba
\label{eq6}
\epsilon_{g,q}&=&\nu_{g,q} \int \frac{d^3p}{(2\pi)^3} \omega_{g,q} f_{g,q}\nn
&=& \pm \frac{3 \nu_{g,q} T^4}{ \pi^2} PolyLog[4, \pm z_{g,q}]\nn
&& \pm \frac{T^4 \nu_{g,q}}{\pi^2} T  \partial_{T} \ln(z_{g,q}) PolyLog[3, \pm z_{g,q}].
\ea
The first  term in the right-hand side  of Eq. (\ref{eq6}) is nothing but the $3 P_{g,q}$, while second term  leads to non-vanishing interaction measure in hot QCD. The entropy density and enthalpy can be read off from the expressions of $\epsilon$ and $P$ using well known thermodynamic relations.
The energy density and enthalpy density per particle can easily  obtained employing results from Eqs.(\ref{eq3}-\ref{eq6}).

\subsubsection*{Charge renormalization and effective coupling}
In contrast to the effective mass models where the effective mass is motivated from the mass renormalization  in the hot QCD medium, 
 the EQPM is based on the charge renormalization in high temperature QCD. 
\begin{figure}[h]
\includegraphics[scale=0.32]{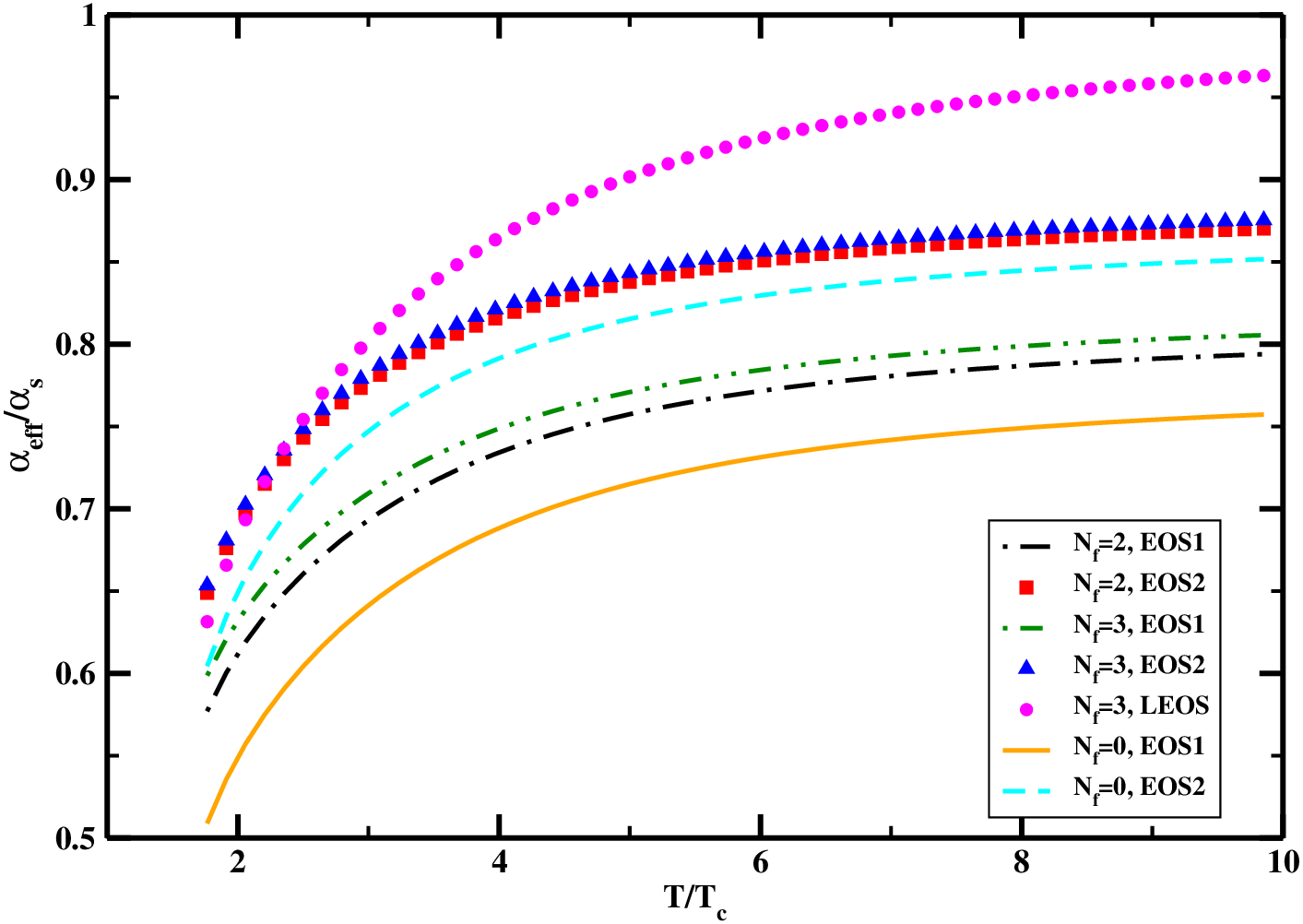}
\caption{(Color online)Effective coupling constant using various EOS as a function of $T/T_c$.}
\label{alphas}
\end{figure} 

 To investigate how the quasi partonic charges modify in the presence of hot QCD medium, 
 we consider the expression for the Debye mass derived in semi-classical transport theory~\cite{dmass1, dm_rev1, dm_rev2} as,
 \ba
 \label{dm}
 m_D^2&=& 4 \pi \alpha_{s}(T) \bigg(-2 N_c \int \frac{d^3 p}{(2 \pi)^3} \partial_p f_g (\vec{p})\nn
 &+& 2 N_f  \int \frac{d^3 p}{(2 \pi)^3} \partial_p f_q (\vec{p})\bigg),
 \ea
 where, $\alpha_{s}(T)$ is the QCD running coupling constant at finite temperature~\cite{qcd_coupling}.

 After performing the momentum integral after substituting the quasi-parton distribution function from  Eq. (\ref{eq1}) to Eq. (\ref{dm}), we obtain,
 \ba
 \label{dm1}
 m_D^2&=&4 \pi \alpha_{s}(T) T^2  \bigg( \frac{2 N_c}{\pi^2} PolyLog[2,z_g]\nn&-&\frac{2 N_f}{\pi^2} PolyLog[2,-z_q]\bigg). 
 \ea
 The Debye mass here reduces to the leading order HTL expression in the limit $z_{g,q}=1$ (ideal EoS: noninteracting of ultra relativistic quarks and gluons),
 \be
 m_D^2(HTL)= \alpha_{s}(T) \ T^2 (\frac{N_c}{3}+\frac{N_f}{6}).
\ee
Eq. \ref{dm1} can be rewritten as, 
\ba
m_D^2&=& m_D^2(HTL)\nn &&\times \frac{\frac{2 N_c}{\pi^2} PolyLog[2,z_g]-\frac{2 N_f}{\pi^2} PolyLog[2,-z_q]}{\frac{N_c}{3}+\frac{N_f}{6}}.\nn
\ea
We can now define the effective coupling, $\alpha_{eff} \equiv \alpha_{s}(T) g(z_g,z_q)$, so that the $m_D^2 =4\pi \alpha_{eff} (T)\  T^2 (N_c/3+N_f/6)$.  The function $g(z_g,z_q)$ reads,\\
\be
g(z_g,z_q)= \frac{\frac{2 N_c}{\pi^2} PolyLog[2,z_g]-\frac{2 N_f}{\pi^2} PolyLog[2,-z_q]}{\frac{N_c}{3}+\frac{N_f}{6}}
\ee

Notably, the EQPM employed here has been remarkably useful in understanding the bulk and the transport properties of the QGP in heavy-ion collisions~\cite{chandra_eta, chandra_etazeta, chandra_dilep, chandra_hq1,chandra_hq2}.

The behavior of the ratio $\alpha_{eff}/\alpha_{s}\equiv g(z_g, z_q)$ as a function of temperature ($T/T_c$) for various EOSs  is depicted in Fig. \ref{alphas}. The flavor dependence is also shown in Fig. \ref{alphas}.
Clearly the ratio will approach to its value with ideal EOS ($z_{g,q}\rightarrow 1$) which is unity, only asymptotically. The EOS dependence can clearly be visualized  from the temperature dependence of the 
relative coupling in Fig. \ref{alphas}. For example the LEOS result is closest to the running $\alpha_s$ among all the cases. Similiary, other EOS dependent predictions can also be explicated.

There are only two free  functions  ($z_g$, and $z_q$)  in the EQPM employed here which depend on the chosen  EOS. In the case of EOS1 and EOS2 employed in the present case, these functions are obtained in~\cite{chandra_quasi1} and are continuos functions 
of $T/T_c$. On the other hand, for LEOS they are defined in terms of eight parameters obtained in Ref.~\cite{chandra_quasi2} (See Table I of Ref.~\cite{chandra_quasi2}). Apart from that effective coupling mentioned above 
depends on the  them and the QCD running coupling constant $g(T)$, that explicitly depends upon how we fix the QCD renormalization scale at finite temperature and up to what order we define $g(T)$.  Henceforth, these are the three quantities that need to be supplied throughout the 
analysis  here.

\subsection{Thermal Relaxation times}

In order to estimate the relaxation times of particles due to their mutual interactions we start with the Boltzmann transport
equation for an out of equilibrium system that describes the binary elastic process $p_{k}+p_{l}\rightarrow p'_{k}+p'_{l}$,
\begin{equation}
 \frac{df_{k}(x,p_{k})}{dt}=-C[f_{k}]~.
 \label{trns-0}
\end{equation}

Here $f_{k}$ is the single particle distribution function for the $k^{th}$ species in a multicomponent
system, that depends upon the particle 4-momentum $p_{k}$ and 4-space-time coordinates $x$. $C[f_{k}]$ 
denotes the collision term that quantifies the rate of change of $f_{k}$ given in the following manner ~\cite{AMY2},

\begin{eqnarray}
 C[f_{k}]=&& \frac{1}{2}\nu_{l}\sum_{l=1}^{N} \frac{1}{2\omega_{k}} \int d\Gamma_{p^{}_{l}} d\Gamma_{p'_{k}} d\Gamma_{p'_{l}} (2\pi)^4 \nn\times
 &&\delta^{4}(p_{k}+p_{l}-p'_{k}-p'_{l}) \langle|M_{k+l\rightarrow k+l}|^{2}\rangle \nonumber \\
 &&\times [f_{k}(p_{k})f_{l}(p_{l})\{1\pm f_{k}(p'_{k})\}\{1\pm f_{l}(p'_{l})\}\nn&&-
        f_{k}(p'_{k}) f_{l}(p'_{l}) \{1\pm f_{k}(p_{k})\}\{1\pm f_{l}(p_{l})\}]~ 
        \label{coll11}\\
        && k=1,2,......,N\nonumber.
\end{eqnarray}

The phase space factor is expressed by the notation $d\Gamma_{p_{i}}=\frac{d^3 \vec {p_{i}} }{(2\pi)^3 2\omega_i}$,
as $\omega_{k}$ is the energy of the scattered particle which is of $k^{th}$ species. 
The overall $\frac{1}{2}$ factor is appearing due to the symmetry in order to compensate for the double counting of
final states that occurs by interchanging $p'_{k}$ and $p'_{l}$.
$\nu_{l}$ is the degeneracy of $2^{nd}$ particle that belongs to $l^{th}$ species.
It is considered next that the out of equilibrium distribution function of the $1^{st}$ particle, which
is being scattered is given by,
\begin{eqnarray}
 f_{k}=f_{k}^0 +\delta f_{k}=f_{k}^0 + f_{k}^0 (1\pm f_{k}^0) \phi_{k}~,
 \label{phi11}
 \end{eqnarray}
where the non-equilibrium part $\delta f_{k}$ of the distribution function is quantified by the deviation
function $\phi_k$. The collision term can now be expressed as the distribution deviation over the relaxation 
time $\tau_k$, which is needed by the out of equilibrium distribution function to restore its 
equilibrium value,

\begin{equation}
C[f_{k}]=\frac{\delta f_{k}}{\tau_k}=\frac{f_{k}^0 (1\pm f_{k}^0) \phi_{k}}{\tau_k}~.
\label{coll2}
\end{equation}

Putting (\ref{phi11}) into the right hand side of (\ref{coll11}) by keeping the distribution functions of 
the particles other than the scattered one vanishingly close to equilibrium and comparing with (\ref{coll2}), the relaxation time 
finally becomes as the inverse of the reaction rate $\Gamma_k$ of the respective processes ~\cite{Zhang},
\begin{eqnarray}
&&\tau_{k}^{-1}\equiv\Gamma_{k}\nonumber\\
&&=\frac{\nu_{l}}{2} \frac{1}{2\omega_{k}} \int d\Gamma_{p_{l}} d\Gamma_{p'_{k}} d\Gamma_{p'_{l}} (2\pi)^4\delta^{4}(p_{k}+p_{l}-p'_{k}-p'_{l})\nonumber\\
&&\times \langle|M_{k+l\rightarrow k+l}|^{2}\rangle \frac{f_{l}^{'0} (1\pm f_{k}^{'0}) (1\pm f_{l}^{0})}{(1\pm f_{k}^{0})}.
\end{eqnarray}
Clearly the distribution function of final state particles are given by primed notation.

Simplifying $\tau_{k}$ utilizing the $\delta$-function we finally obtain $\tau_k$ in the center of momentum 
frame of particle interaction as,

\begin{eqnarray}
 &&\tau_k^{-1}=\Gamma_{k}=\nonumber\\
 && \nu_{l}\int \frac{d^3 \vec{p_{l}}}{(2\pi)^3} d(\cos\theta) \frac{d\sigma}{d(\cos\theta)} 
 \frac{f_{l}^{'0} (1\pm f_{k}^{'0}) (1\pm f_{l}^{0})}{(1\pm f_{k}^{0})}~,
\end{eqnarray}
where $\theta$ is the scattering angle in the center of momentum frame and $\sigma$ is the interaction
cross section for the respective scattering processes. Now in terms of the Mandelstam variables $s,t$ 
and $u$ the expression for $\tau_{k}$ can be reduced simply as,

\begin{equation}
 \tau_{k}^{-1}=\Gamma_{k}=\nu_{l}\int \frac{d^3 \vec{p_{l}}}{(2\pi)^3} dt \frac{d\sigma}{dt} 
 \frac{f_{l}^{'0} (1\pm f_{k}^{'0}) (1\pm f_{l}^{0})}{(1\pm f_{k}^{0})}~.
\end{equation}

The differential cross section relates the scattering amplitudes as $\frac{d\sigma}{dt}=\frac{\langle|M|^2\rangle}{16\pi s^2}$.
The quark-gluon scattering amplitudes for $2\rightarrow2$ processes are taken from ~\cite{Combridge}, 
that are averaged over the spin and color degrees of freedom of the initial states and summed over the  
final states.
 
Now in order to take into account the small-angle scattering scenario that results into divergent contributions 
from $t$-channel diagrams of QCD interactions, a transport weigh factor $(1-\cos\theta)=\frac{2tu}{s^2}$ 
have been introduced in the interaction rate ~\cite{Thoma,Hosoya}. Furthermore considering the momentum transfer
$q=|\vec{p_k}-\vec{p'_{k}}|=|\vec{p_{l}}-\vec{p'_{l}}|$ is not too large we can make following assumptions,
$f_{k}^{0}\cong f_{k}^{'0}$ and $f_{l}^{0}\cong f_{l}^{'0}$ ~\cite{Thoma} to finally obtain,

\begin{equation}
 \tau_k^{-1}=\Gamma_{k}=\nu_{l}\int \frac{d^3 \vec{p_{l}}}{(2\pi)^3} f_{l}^{0}(1\pm f_{l}^{0})
 \int dt \frac{d\sigma}{dt} \frac{2tu}{s^2}~.
 \label{tau}
\end{equation}

In the integration involving $t$-channel diagrams from where the infrared logarithmic singularity appears,
the limit of integration is restricted from $-s$ to $-k^2$ in order to avoid those divergent results 
using the cut-off $k^2=g^2 T^2$ as infrared regulator. Here $g^2=4\pi\alpha_s$ with $\alpha_s$ being 
the coupling constant of strong interaction as already mentioned in section-A. 

Now in the QGP medium the quark and gluon interaction rates result from the following interactions respectively,
\begin{equation}
 \Gamma_g=\Gamma_{gg}+\Gamma_{gq}~,~~~~~~
 \Gamma_q=\Gamma_{qg}+\Gamma_{qq}~,
\end{equation}
where $\Gamma_{gg}$, $\Gamma_{gq}, \Gamma_{qg}$ and $\Gamma_{qq}$ are the interaction rates between  
gluon-gluon, gluon-quark, quark-gluon and quark-quark respectively.

Finally after pursuing the angular integration in (\ref{tau}) we are left with the thermal relaxation times of the 
quark and gluon components in a QGP system in the following way,

\begin{eqnarray}
 \tau_g^{-1}=&& \{\nu_{g}\int \frac{d^3 \vec{p_{l}}}{(2\pi)^3} f_{g}^{0}(1+f_{g}^{0})\} \nonumber \\
      \times &&[\frac{9g^4}{16\pi\langle s \rangle_{gg}}\{ln\frac{\langle s \rangle_{gg}}{k^2}-1.267 \}] \nonumber \\
            +&& \{\nu_{q}\int \frac{d^3 \vec{p_{l}}}{(2\pi)^3} f_{q}^{0}(1-f_{q}^{0})\} \nonumber \\
      \times &&[\frac{g^4}{4\pi\langle s \rangle_{gq}}\{ln\frac{\langle s \rangle_{gq}}{k^2}-1.287 \}],
 \label{taug}
 \end{eqnarray}
 
 \begin{eqnarray}
 \tau_q^{-1}=&& \{\nu_{g}\int \frac{d^3 \vec{p_{l}}}{(2\pi)^3} f_{g}^{0}(1+f_{g}^{0})\} \nonumber \\
      \times &&[\frac{g^4}{4\pi\langle s \rangle_{qg}}\{ln\frac{\langle s \rangle_{qg}}{k^2}-1.287 \}] \nonumber \\
            +&& \{\nu_{q}\int \frac{d^3 \vec{p_{l}}}{(2\pi)^3} f_{q}^{0}(1-f_{q}^{0})\} \nonumber \\
      \times &&[\frac{g^4}{9\pi\langle s \rangle_{qq}}\{ln\frac{\langle s \rangle_{qq}}{k^2}-1.417 \}],
 \label{tauq}
\end{eqnarray}

where $\langle s \rangle_{kl}=2\langle p_{k} \rangle \langle p_{l} \rangle$ is the thermal average value of $s$ with
$\langle p_k \rangle =\frac{\int \frac{d^3 \vec{p_{k}}}{(2\pi)^3} |\vec{p_{k}}| f_{k}^{0}}{\int \frac{d^3 \vec{p_{k}}}{(2\pi)^3} f_{k}^{0}}$.
Clearly in order to account for a hot QCD medium the quasiparticle effects must be invoked in the expressions
of these thermal relaxation times obtained far. As discussed in section-A, the distribution functions of quarks
and gluons and the coupling $g$ will carry the quasiparticle descriptions accordingly. Since the cut-off parameter
$k$ also depends upon $g$ and the thermal average of $s$ includes $f^{0}_{g/q}$, they will reflect the hot QCD equation 
of state effect as well. Following the definition of equilibrium distribution function of quarks and gluons from Eq.(\ref{eq1}), 
within the quasiparticle framework, the thermal averages of gluon and quark momenta respectively  are obtained as,  
\begin{eqnarray}
 \langle p_g \rangle =3T\frac{PolyLog[4,z_{g}]}{PolyLog[3,z_{g}]}~,\\
 \langle p_q \rangle =3T\frac{PolyLog[4,-z_{q}]}{PolyLog[3,-z_{q}]}~.
\end{eqnarray}

\subsection{Electrical conductivity}
In this work, we have adopted the kinetic theory approach for evaluating the analytical expression 
of electrical conductivity, based on solving the relativistic transport equation for a charged QGP system.

Before proceeding for the solution of transport equation, we  introduce here some  of the thermodynamic
quantities needed for developing the required  framework. 
We start with particle 4-flow for the $k^{th}$ species of particle in a multicomponent system ~\cite{Degroot},
\begin{equation}
N_{k}^{\mu}(x)=\int \frac{d^{3}\vec{p_{k}}}{(2\pi)^3 p_{k}^0} p_{k}^{\mu}f_{k}(x,p_{k})~. 
\end{equation}
Next the total particle 4-flow and the energy momentum tensor of the system are defined respectively as the following,
\begin{eqnarray}
 N^{\mu}(x)=&&\sum_{k=1}^{N}N_{k}^{\mu}(x)=\sum_{k=1}^{N}\int \frac{d^{3}\vec{p_{k}}}{(2\pi)^3 p_{k}^0} p_{k}^{\mu}f_{k}(x,p_{k})~,\\
 T^{\mu\nu}(x)=&&\sum_{k=1}^{N}\int \frac{d^{3}\vec{p_{k}}}{(2\pi)^3 p_{k}^0} p_{k}^{\mu}p_{k}^{\nu}f_{k}(x,p)~.
\end{eqnarray}
With the help of the above quantities we define the diffusion flow of the $k^{th}$ component as ~\cite{Degroot},
\begin{equation}
 I_{k}^{\mu}=N_{k}^{\mu}-x_{k}N^{\mu}~,
\end{equation}
where $x_{k}=\frac{n_{k}}{n}$ is the particle fraction corresponding to $k^{th}$ species, $n_{k}$
and $n$ are the particle number density for $k^{th}$ species and total particle number density of the 
multicomponent system 
respectively, which are related by $n=\sum_{k=1}^{N}n_{k}$. We can readily notice $\sum_{k=1}^{N}I_{k}=0$, 
i,e. sum of the diffusion flows vanishes.

The total electric current density of such a system is given by ~\cite{Greif},
\begin{equation}
 J^{\mu}(x)=\sum_{k=1}^{N}q_{k}I_{k}^{\mu}=\sum_{k=1}^{N-1}(q_{k}-q_{N})I_{k}^{\mu}~,
 \label{current}
\end{equation}
where $q_{k}$ is the electric charge associated with the $k^{th}$ species.

A realistic description of non-equilibrium phenomena in relativistic systems must take reactive
processes into account which incorporates all kinds of inelastic collisions beside elastic ones.
In such a case the system must include a number of conserved quantum numbers and the diffusion
flow in such situations will become,
\begin{eqnarray}
I_{a}^{\mu}=&&\sum_{k=1}^{N}q_{ak}I_{k},~~~~~~~~~~~~~~~~[a=1,2,..........N']\\
           =&&\sum_{k=1}^{N}q_{ak}\{N_{k}^{\mu}-x_{k}N^{\mu}\}~.
\label{diff}
\end{eqnarray}
Here $a$ stands for the index of conserved quantum number and $q_{ak}$ is the $a^{th}$ conserved 
quantum number associated with $k^{th}$ component. Following the prescription we are able to
define the particle number density of the independent components as, $n_{a}=\sum_{k=1}^{N}q_{ak}n_{k}$.

After defining these basic thermodynamic quantities let us present the relativistic transport
equation (\ref{trns-0}) in covariant form with the force term present in it ~\cite{Xu-Greiner},

\begin{equation}
p_{k}^{\mu}\partial_{\mu}f_{k}+q_{k}F^{\alpha\beta}p_{\beta}\frac{\partial f_{k}}{\partial p_{k}^{\alpha}}=-C[f_k]~.
\end{equation}
Here $F^{\mu\nu}=\{-u^{\mu}E^{\nu}+u^{\nu}E^{\mu}\}$ is the electromagnetic field tensor with electric field 
$E^{\mu}$, in the absence of
any magnetic field in the medium. We identify $u^{\mu}$ as hydrodynamic 4-velocity.
Throughout this paper we will use the metric system $g^{\mu\nu}=\{1,-1,-1,-1\}$.

Now using the Chapman-Enskog (CE) method the transport equation is linearized around a local equilibrium 
distribution function $f_{k}^{0}(x,p_{k})$ and finally the CE hierarchy reduces the left hand side of the 
transport equation in terms of $f_{k}^{0}$. The collision term is simplified using (\ref{coll2}) giving
rise to,

\begin{equation}
p_{k}^{\mu}\partial_{\mu}f_{k}^{0}+\frac{1}{T}f_{k}^0(1\pm f_{k}^0)q_{k}E_{\mu}p_{k}^{\mu}=
-\frac{\omega_{k}}{\tau_{k}}f_{k}^0(1\pm f_{k}^0)\phi_{k}~.
\label{trns-1}
\end{equation}
To proceed further the distribution functions of constituent particles is needed to be provided in covariant notations.
In a comoving frame and involving the quasiparticle description discussed in section-A, it can be given in the following way,
\begin{equation}
 f_{k}^{0}(x,p_{k})=
 \frac{z_{k}exp[-\frac{p_{k}^{\mu}u_{\mu}}{T}+\frac{\mu_{k}}{T}]}{1\mp z_{k}exp[-\frac{p_{k}^{\mu}u_{\mu}}{T}+\frac{\mu_{k}}{T}]}~,
\end{equation}
where we have introduced $\omega_{k}$ as the energy per particle of the $k^{th}$ species and $\mu_{k}$ is
the chemical potential for the same. Within the quasiparticle framework, for quarks
and gluons $\omega_k$ is defined by Eq.(\ref{epp}). 

In order to retrieve the transport equation in terms of the thermodynamic forces, the first term on the left hand
side of Eq.(\ref{trns-1})
is needed to be reduced by decomposing the derivative over the distribution function into a time-like and a space-like
part as $\partial^{\mu}=u^{\mu}D+\nabla^{\mu}$, with the covariant time derivative $D=u^{\mu}\partial_{\mu}$ 
and the spatial gradient $\nabla_{\mu}=\Delta_{\mu\nu}\partial^{\nu}$, expressed in terms of hydrodynamic 4-velocity
$u^{\mu}$ and projection operator $\Delta_{\mu\nu}=g_{\mu\nu}-u_{\mu}u_{\nu}$. Whence the spatial gradients over
velocity, temperature and chemical potentials directly link with the viscous flow, heat flow and the diffusion flow
of the fluid respectively, the time derivatives are needed to be eliminated using a number of thermodynamic identities
so that they contribute in the expressions of the thermodynamic forces as well. The time derivative
over particle number density and the time derivative over energy per particle however follow the equation of 
continuity and equation of energy as in the case of a system without the influence of electric field,
\begin{eqnarray}
Dn_{k}=&&-n_{k}\partial\cdot u ~,
\label{numberid}
\\
\sum_{k=1}^{N}x_{k}D\omega_{k}=&&-\frac{\sum_{k=1}^N P_{k}}{\sum_{k=1}^N n_{k}}\partial \cdot u ~,
\label{energyid}
\end{eqnarray}
where $P_k$ is the partial pressure attributed to $k^{th}$ species.
But the equation of motion in the presence of the electric field will be different from the one without electric field.
In a multicomponent system in the presence of an electric field the equation of motion takes the follow form,
\begin{equation}
 Du^{\mu}=\frac{\nabla^{\mu} P}{\sum_{k=1}^{N} n_{k} h_{k}}+\frac{\sum_{k=1}^N q_{k}n_{k}}{\sum_{k=1}^N h_{k}n_{k}}E^{\mu}~.
 \label{velid}
 \end{equation}
Clearly even the pressure gradient is zero, the Lorentz force acting on the particle produces non-zero acceleration.
By utilizing these identities and 
retaining the thermodynamic forces involving thermal and diffusion terms only, (shear and 
bulk viscous part not considered in this work), the transport equation becomes,

\begin{equation}
\frac{1}{T} [p_{k}^{\nu}\{(p_{k}.u)-h_{k}\}X_{qk}+p_{k}^{\nu}\sum_{a=1}^{N'-1}(q_{ak}-x_{a})X_{a\nu}]=-\frac{\omega_{k}}{\tau_{k}}\phi_{k}~,
\label{trns-2}
\end{equation}

where $X_{q\mu}$ and $X_{a\mu}$ are the thermal and diffusion forces respectively given by, 

\begin{eqnarray}
X_{q\mu}=&&[\frac{\nabla_{\mu}T}{T}-\frac{\nabla_{\mu}P}{nh}]+[-\frac{1}{h}\sum_{k=1}^{N}x_{k}q_{k}E_{\mu}]~,
\label{forceth}\\
X_{k\mu}=&&[(\nabla_{\mu}\mu_{a})_{P,T}-\frac{h_{k}}{nh}\nabla_{\mu}P]+\nonumber\\
         &&[q_{k}-q_{N}-\frac{h_{k}-h_{N}}{h}\sum_{l=1}^{N}x_{l}q_{l}]E_{\mu}~.
\label{forcedif}
\end{eqnarray}
The detail of the computation  in offered  in Appendix-A.

We identify $h_{k}$ and $h$ as the enthalpy per particle for species $k$ and for total system
respectively and $(\nabla_{\mu}\mu_{a})_{P,T}=\sum_{b=1}^{N'-1}\{\frac{\partial \mu_a}{\partial x_b}\}_{P,T,\{x_{a}\}} \nabla_{\mu}x_b$. 
Here $x_a$ and $\mu_a$ are the particle fraction and chemical potential associated with $a^{th}$ quantum number respectively. 
Clearly in the expressions of thermal and diffusion driving forces, terms proportional to electric field give 
rise to electrical conductivity.
Now in order to be a solution of this equation the deviation function $\phi_{k}$ must be a linear
combination of the thermodynamic forces,
\begin{equation}
\phi_{k}=B_{k\mu}X_{q}^{\mu}+\frac{1}{T}\sum_{a=1}^{N'-1}B_{ak}^{\mu}X_{a\mu}~,
\label{phi}
\end{equation}
with, $B_{k}^{\mu}=B_{k}\langle\Pi_{k}^{\mu}\rangle$ and $B_{ak}^{\mu}=B_{ak}\langle\Pi_{k}^{\mu}\rangle$
where $\langle\Pi_{k}^{\mu}\rangle=(\Pi_{k})_{\nu}\Delta^{\mu\nu}$ and $\Pi_{k}=p_{k}/T$.

Putting (\ref{phi}) into the right hand side of (\ref{trns-2}) and comparing both sides of (\ref{trns-2}) (noting thermodynamic forces
are independent) we finally obtain,
\begin{eqnarray}
B_{k}^{\mu}=\langle\Pi_{k}^{\mu}\rangle\frac{\omega_{k}-h_{k}}{\{-\frac{\omega_{k}}{\tau_k}\}}~,~~~~
B_{ak}^{\mu}=\langle\Pi_{k}^{\mu}\rangle\frac{q_{ak}-x_{a}}{\{-\frac{\omega_{k}}{\tau_k T}\}}~,
\label{B}
\end{eqnarray}
from which the complete structure of $\phi_k$ can be obtained.
Now going back to equation (\ref{diff}) we notice for equilibrium distribution function $f_{k}^{0}$
the $I_{a}^{\mu}$ clearly vanishes, while with $f_{k}=f_k^0(1\pm f_k^0)\phi_k$ it gives a finite
diffusion flow as following,

\begin{equation}
I_{a}^{\mu}=\sum_{k=1}^{N}(q_{ak}-x_{a})\int \frac{d^{3}\vec{p_{k}}}{(2\pi)^3 p_{k}^0} p_{k}^{\mu} f_k^0(1\pm f_k^0)\phi_k~. 
\label{diff-2}
\end{equation}
Putting the value of $\phi_{k}$ from (\ref{phi}) with the help of Eqs. (\ref{B}) into (\ref{diff-2})
we get the linear law of diffusion flow,
\begin{equation}
I_{a}^{\mu}=l_{aq}X_{q}^{\mu}+\sum_{b=1}^{N'-1}l_{ab}X_{b}^{\mu}~, a=1,....,(N'-1)~.
\label{diff-3}
\end{equation}
where the coefficients are now expressed in terms of the relaxation time $\tau$,
\begin{eqnarray}
l_{aq}&=&\sum_{k=1}^{N}(q_{ak}-x_{a}) \frac{1}{T}\int \frac{d^{3}\vec{p_{k}}}{(2\pi)^3}f_k^0(1\pm f_k^0)\tau_{k} \nn
&&\times(\omega_k-h_k)~,
\end{eqnarray}
\begin{eqnarray}
l_{ab}=\sum_{k=1}^{N}(q_{ak}-x_{a})(q_{bk}-x_{b})\frac{1}{T}\int \frac{d^{3}\vec{p_{k}}}{(2\pi)^3}f_k^0(1\pm f_k^0)\tau_{k}.  
\end{eqnarray}

Now substituting the expression of diffusion flow into Eq. (\ref{current}), and pertaining the terms
proportional to electric field only 
we finally reach the expression for the electric current density,

\begin{eqnarray}
 J^{\mu}&=&\sum_{k=1}^{N-1}(q_k-q_N)[\sum_{l=1}^{N-1}l_{kl}\{q_l-q_N-\frac{h_l-h_N}{h}\sum_{n=1}^{N}x_{n}q_{n}\}\nn&&-
 \frac{l_{kq}}{h}\sum_{n=1}^{N}x_{n}q_{n}]E^{\mu}~.
 \label{J1}
\end{eqnarray}

We also know the current density relates with the electric field by the linear relation via the electrical conductivity as,
\begin{equation}
 J^{\mu}=\sigma_{el}E^{\mu}~.
 \label{J2}
\end{equation}
By comparing (\ref{J1}) and (\ref{J2}) we finally obtain the detailed expression of electrical conductivity in the following manner,
\begin{eqnarray}
 \sigma_{el}=&&\sum_{k=1}^{N-1}(q_k-q_N) [\sum_{l=1}^{N-1}l_{kl}\{q_l-q_N-  \nonumber\\
             &&\frac{h_l-h_N}{h}\sum_{n=1}^{N}x_{n}q_{n}\}- \frac{l_{kq}}{h}\sum_{n=1}^{N}x_{n}q_{n}]~.
 \label{sigma}
\end{eqnarray}
Now for a quark-gluon system the expression of the electric conductivity boils down to,
\begin{equation}
 \sigma_{el}=q_{q}^2 \frac{l_{11}h_{g}-l_{1q}x_{q}}{h}.
 \label{sigma_11}
\end{equation}
The subscript $q$ and $g$ stands for quarks and gluons respectively. So finally we are left with the coefficients as,

\begin{eqnarray}
 l_{1q}=&&\frac{1}{T}[-x_q\tau_g\int \frac{d^3p_{g}}{(2\pi)^3} f_{g}^0(1+f_{g}^0) (\omega_g -h_g)\nn
 &&+x_g\tau_q\int \frac{d^3p_{q}}{(2\pi)^3} f_{q}^0(1-f_{q}^0) (\omega_q -h_q)],
 \end{eqnarray}
 \begin{eqnarray}
 l_{11}=&&\frac{1}{T}[x_q^2\tau_g\int \frac{d^3p_{g}}{(2\pi)^3} f_{g}^0(1+f_{g}^0) \nn&&+x_g^2\tau_q\int \frac{d^3p_{q}}{(2\pi)^3} f_{q}^0(1-f_{q}^0) ] ~ .
 \label{l11}
\end{eqnarray}

The $q_q^2=\sum_k \nu_k q_{qk}^2$ is simply the square of the fractional quark charges taking sum over quark degeneracies.
For up, down and strange quarks the fractions quark charge is taken to be $2/3$, $-1/3$ and $-1/3$ respectively.

\subsection{Charge Diffusion}
We recall Eq. (\ref{diff-3}), where the diffusion flow is linearly
expressed in terms of thermal and diffusion driving forces respectively.
The diffusion driving force does not include the terms containing $a=N'$
because diffusion flow vanishes for those values of $a$.
Since presently we are dealing with a quark-gluon plasma which incorporates
binary elastic collisions that conserve particle numbers, in such case
the distinction between the independent particle fractions $x_{a},~ a=1,\cdots, N'$
and the particle fraction of separate components $x_{k},~ k=1,\cdots, N$ vanishes.
So in present situation the diffusion flow rather follow the relation $\sum_{k=1}^{N}I_{k}=0$,
as mentioned earlier. In such scenario the original diffusion driving forces (not containing
the electric field) conjugate to (N-1)
independent diffusion flows is given by ~\cite{Degroot},

\begin{eqnarray}
 X_{k}^{\mu}=&&[(\nabla^{\mu} \mu_{k})_{P,T}-(\nabla^{\mu} \mu_{N})_{P,T}]-\frac{h_k -h_N }{hn}\nabla^{\mu} P,\nn
 && ~~~~~k=1,2,...(N-1)~.
\end{eqnarray}
It is straightforward to  prove  that $(\nabla^{\mu}\mu_{k})_{P,T}=\frac{T}{x_{k}}\nabla^{\mu}x_{k}$.
Thus, in absence of any electric field in $X_{k}^{\mu}$, the flow becomes purely diffusive
that encodes the spacial variation of the fractional particle density corresponding to
different species. Finally, for a two component quark-gluon system at mechanical equilibrium, 
{\it i.e.} at vanishing pressure gradient we obtain the diffusion flow in the following way,
\begin{equation}
 I_{1}^{\mu}=n x_q x_g D_{T}\nabla^{\mu}T+nD\nabla^{\mu}x_q~.
\end{equation}
So we are able to identify the diffusion coefficient as
\begin{equation}
 D=\frac{T l_{11}}{n x_q x_g}~,
 \label{diff-1}
\end{equation}

and the thermal diffusion coefficient as,
\begin{equation}
 D_{T}=\frac{l_{1q}}{nx_{q}x_{g}T}~.
\end{equation}

Taking the value of $l_{11}$ from (\ref{l11}) and incorporating the sum over all flavors and helicities
of the quarks, interacting among themselves and gluons, we finally able to estimate the charge diffusion coefficient
of the system.

\subsection{Charge susceptibility}
In previous Sections, we obtained both the expressions of electrical conductivity and flavor diffusion.
In absence of any electric field, the out of equilibrium particle distribution function relaxes pure 
diffusively, whereas, in the presence of an electric field the restoration of equilibrium is affected by
the electric conductivity. These two quantities are linearly related by Einstein's relation,

\begin{equation}
\sigma_{el}=\chi D~,
\end{equation}
where the proportionality constant is termed as charge susceptibility.
Clearly this term is independent of the relaxation time $\tau$ and particle interactions.
It depends upon the fractional quark charges and thermodynamic parameters describing the system.
In transport theory this quantity is also of significant interest and hence studied in the present work.

\section{Results and Discussions}
\begin{figure}
\includegraphics[scale=0.32]{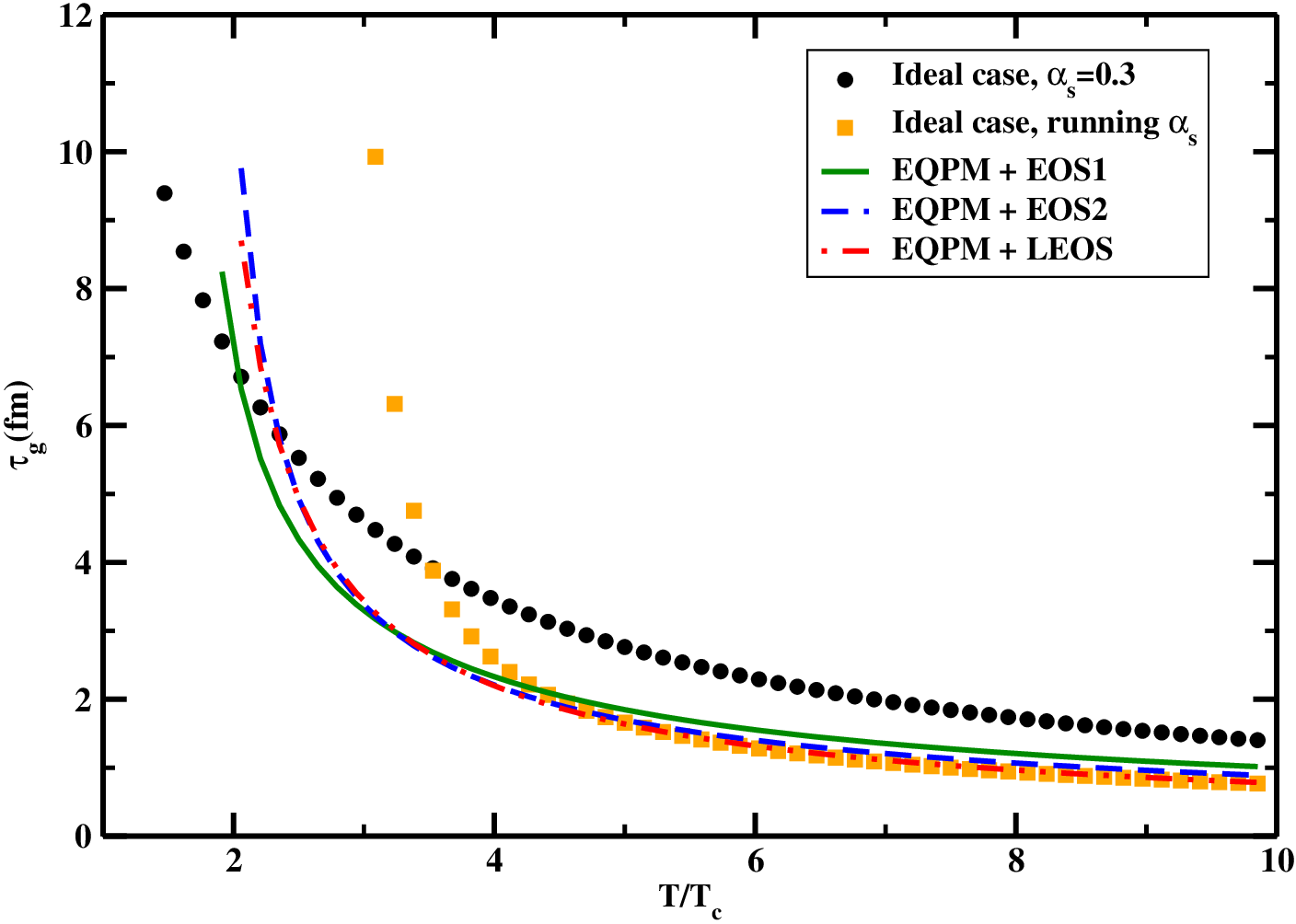}
\caption{(Color online) Temperature dependence of thermal relaxation times for gluons for 3- flavor case .}
\label{taugg}
\end{figure}

\begin{figure}
\includegraphics[scale=0.32]{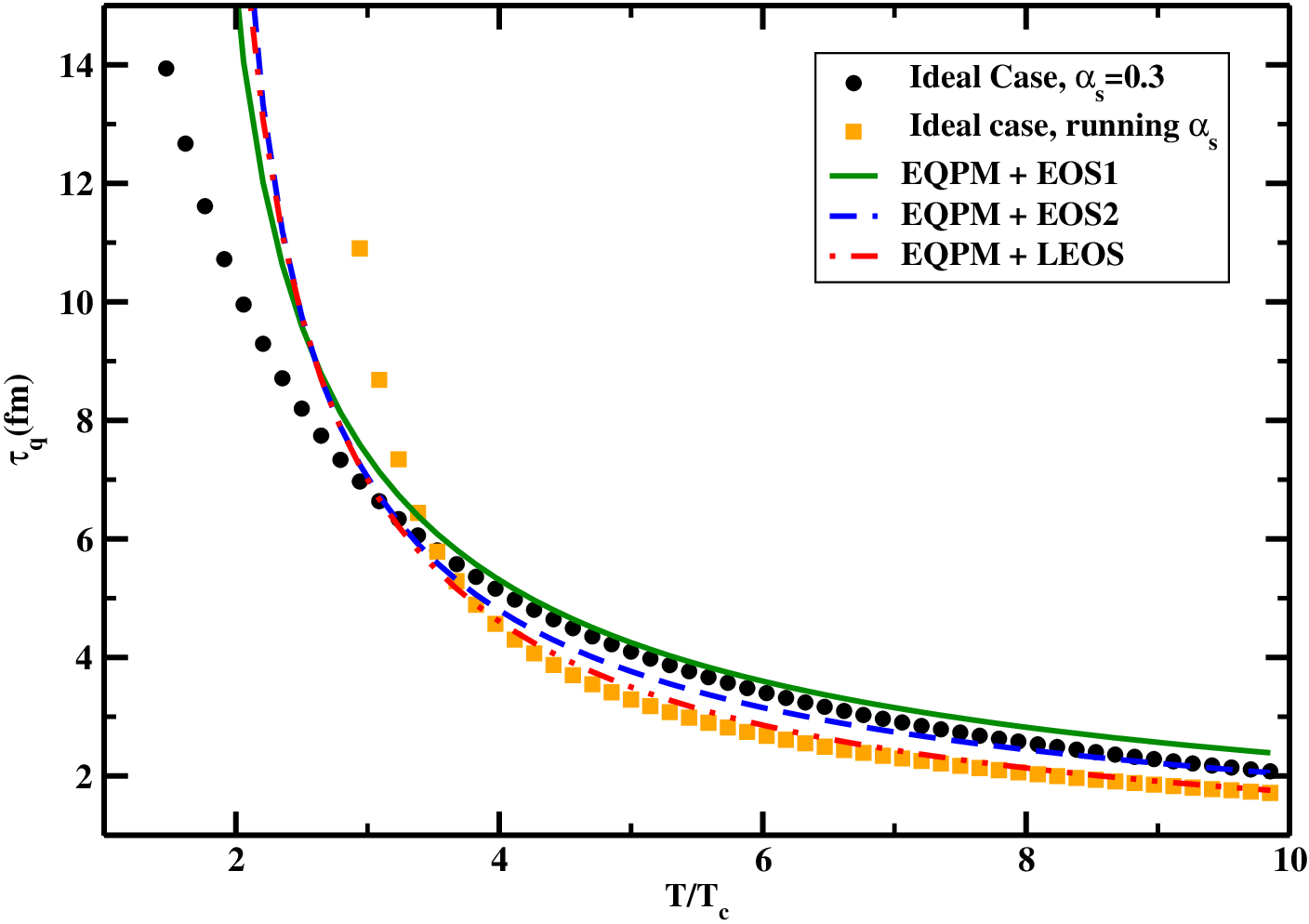}
\caption{(Color online) Temperature dependence of thermal relaxation times for quarks for 3-flavor case.}
\label{taugq}
\end{figure}

\begin{figure}
\includegraphics[scale=0.32]{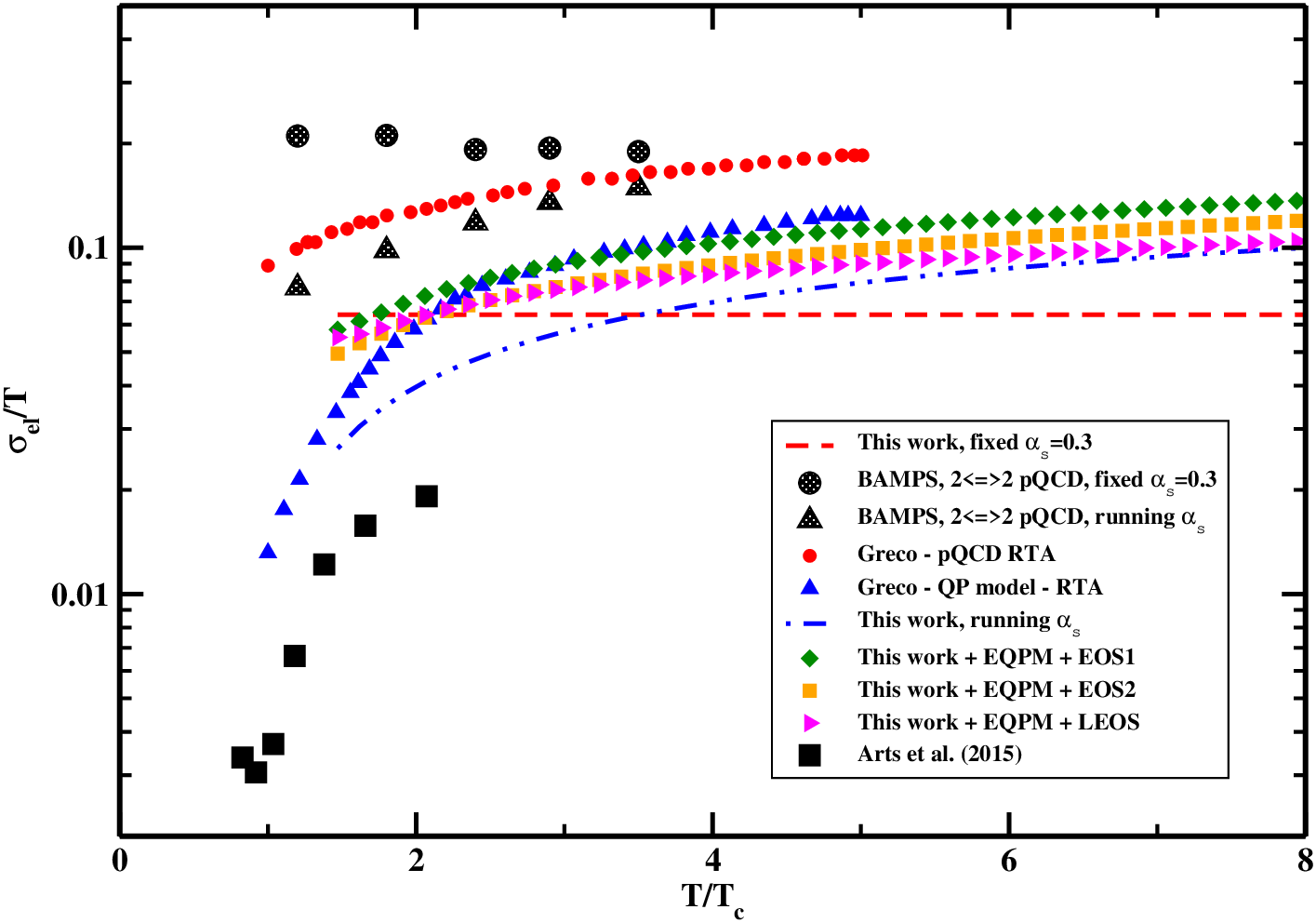}
\caption{(Color online)The electrical conductivity scaled with temperature, $\sigma_{el}/T$ for 3-flavor pQCD cross section as a function of $T/T_c$ employing different EOSs.}
\label{sigmapQCD}
\end{figure}

\begin{figure}
\includegraphics[scale=0.32]{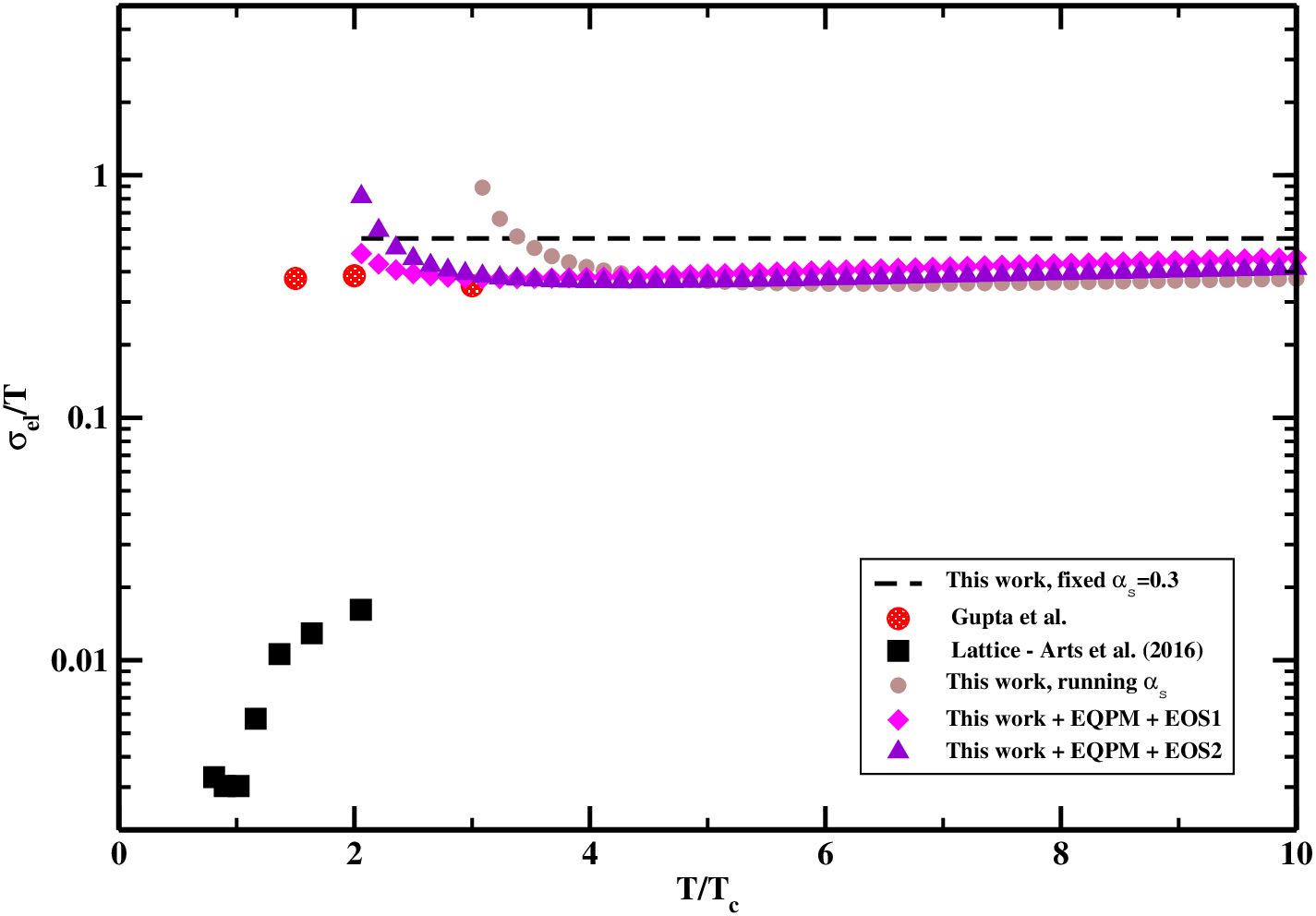}
\caption{(Color online)The electrical conductivity scaled with temperature, $\sigma_{el}/T$ for 2-flavor leading-log cross section as a function of $T/T_c$ employing different EOSs.}
\label{sigma-ll-2}
\end{figure}

\begin{figure}
\includegraphics[scale=0.32]{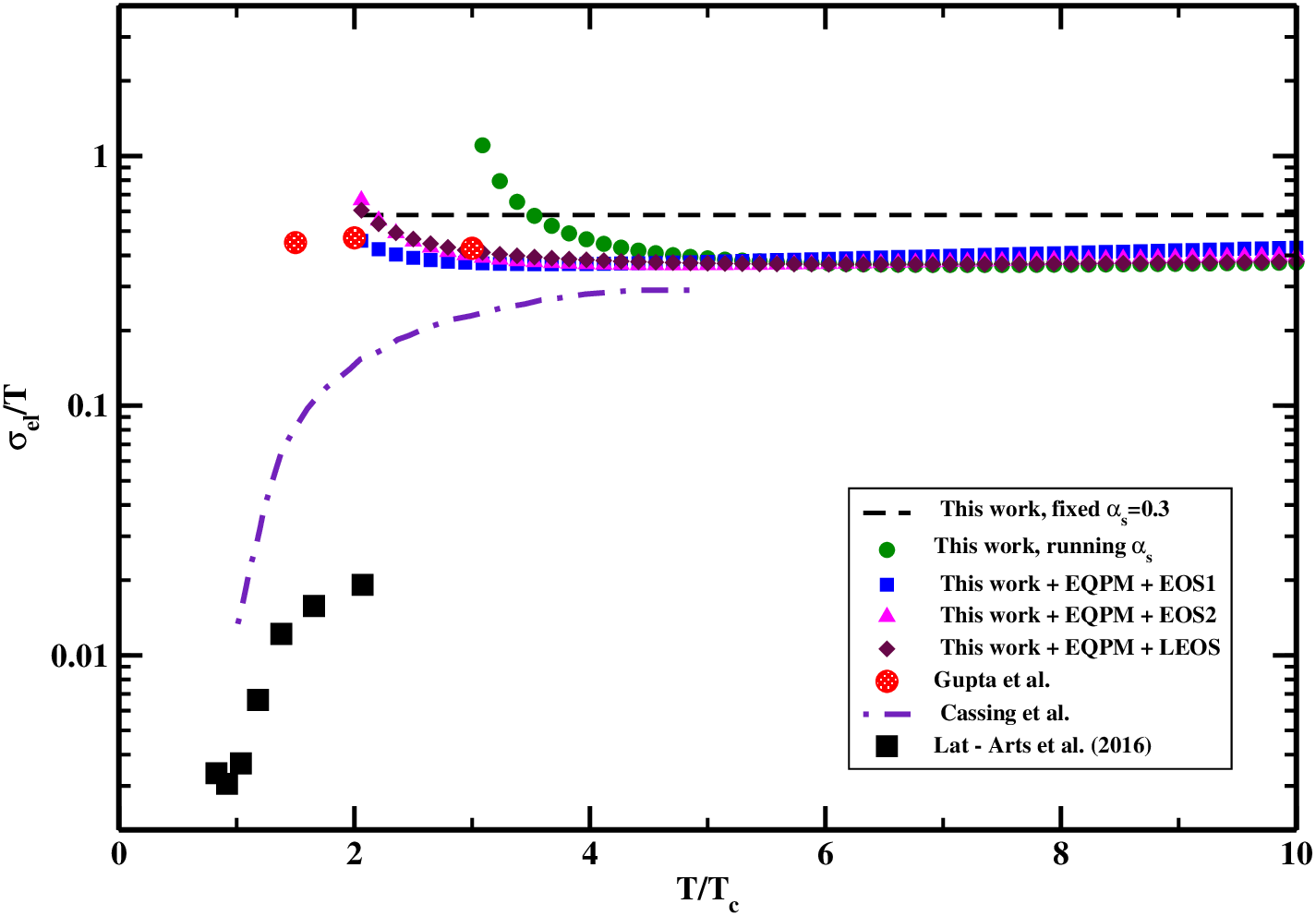}
\caption{(Color online) The electrical conductivity scaled with temperature, $\sigma_{el}/T$ for 3-flavor leading-log cross section as a function of $T/T_c$ employing different EOSs.}
\label{sigma-ll-3}
\end{figure}

\begin{figure}
\includegraphics[scale=0.32]{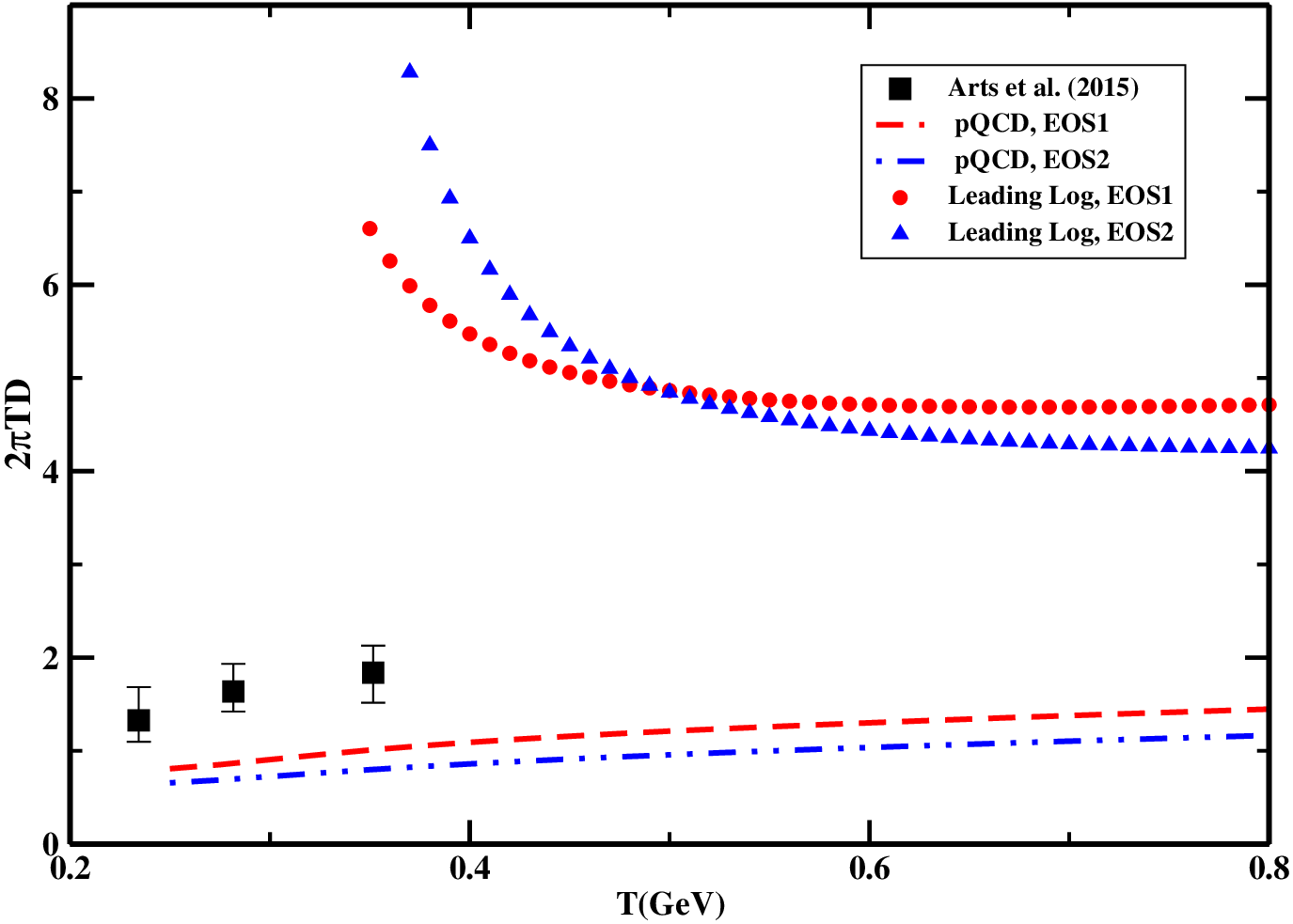}
\caption{(Color online) Scaled charge diffusion coefficient, $2\pi DT$ for 2-flavor using pQCD and leading-log cross sections as a function of $T$ employing various  EOSs.}
\label{Diff2fig}
\end{figure}

\begin{figure}
\includegraphics[scale=0.32]{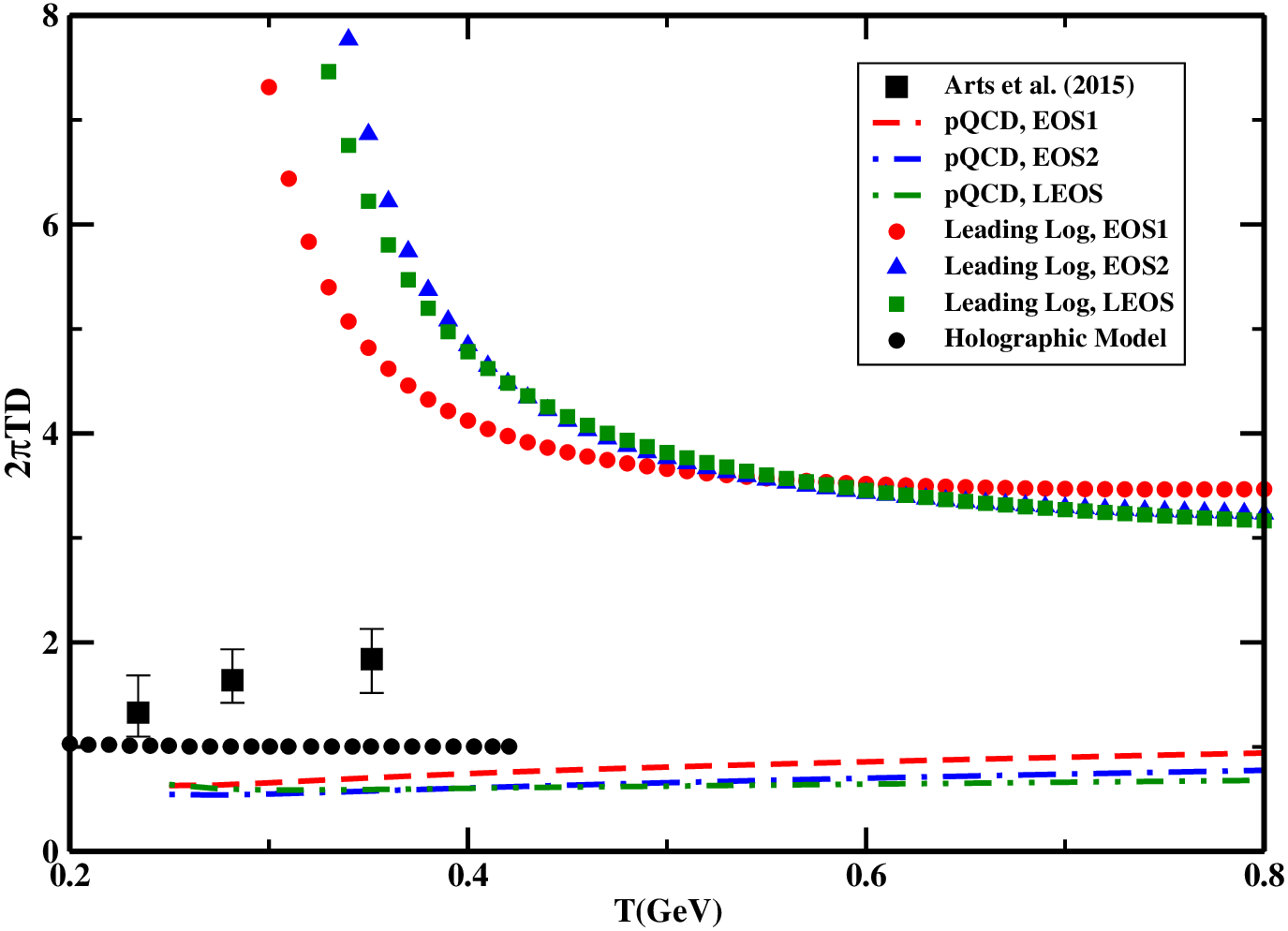}
\caption{(Color online)  Scaled diffusion coefficient, $2\pi DT$ for 3-flavor using pQCD and leading-log cross sections as a function of $T$ employing  various EOSs.}
\label{Diff3fig}
\end{figure}

\begin{figure} 
\includegraphics[scale=0.32]{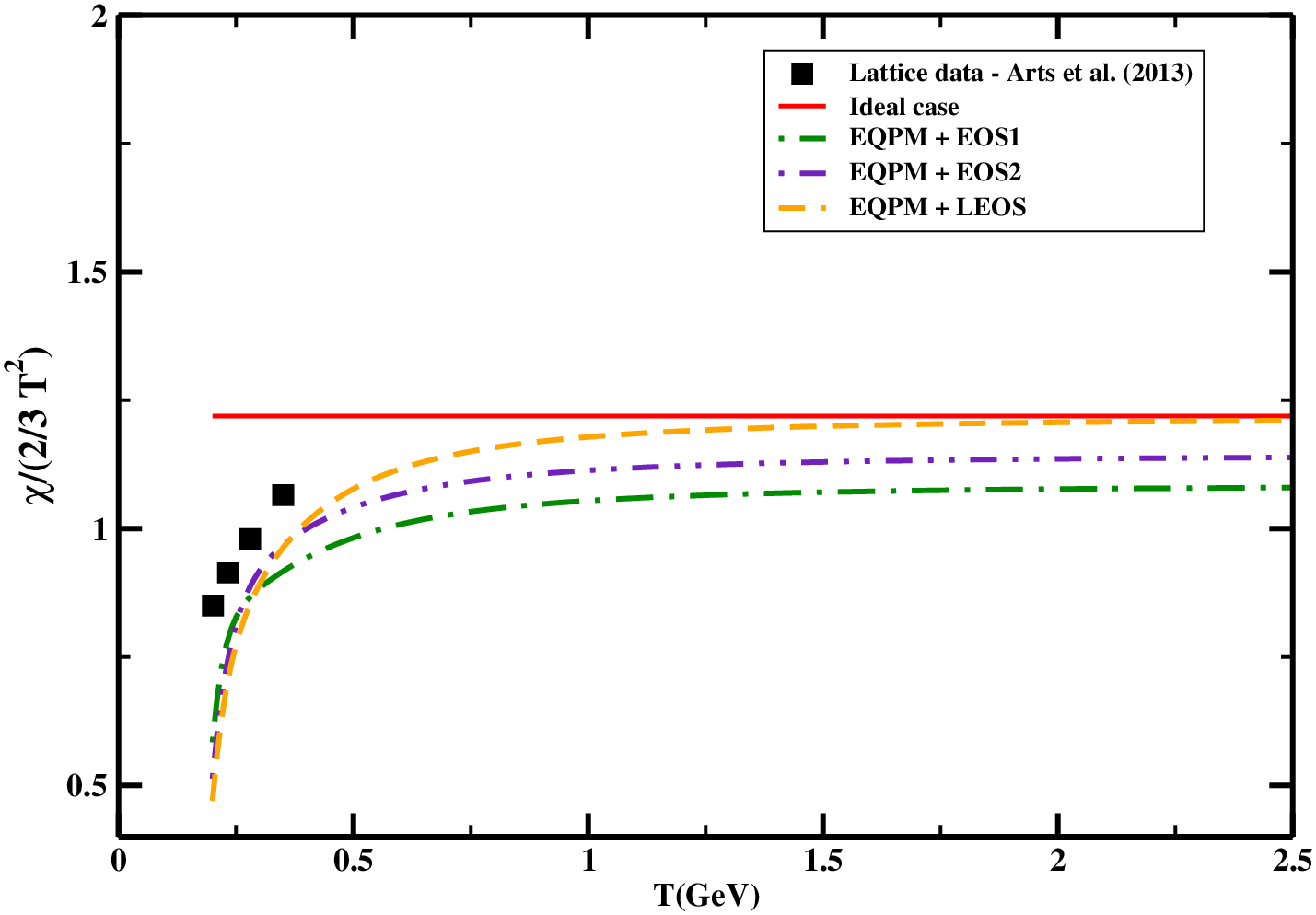}
\caption{(Color online) Susceptibility, $\chi$ for 3-flavor using pQCD and leading-log cross sections as a function of $T$ employing various  EOSs.}
\label{susc}
\end{figure}
In this section,  we initiate our discussions with the temperature dependence of the 
 thermal relaxation times of quarks and gluons considering 3-flavors
of quarks (up, down and strange). Following from Eq.(\ref{taug}) and (\ref{tauq}), $\tau_{g}$ and $\tau_{q}$
have been plotted as a function of $T/T_{c}$ for different $\alpha_s$ in Fig.\ref{taugg} and Fig.\ref{taugq} 
respectively. Both $\tau_g$ and $\tau_q$
exhibit the expected decreasing trend with increasing temperature. 
This observation reveals that  at higher temperature,
the increased interaction rates make the quarks and gluons to restore their equilibrium faster.
The order of magnitude of $\tau_g$ and $\tau_q$ and the fact that $\tau_q$ is larger than $\tau_g$ agree with 
the work in ~\cite{Baym}. The thermal relaxation times for both the cases have been estimated for different values of the QCD 
couplings. 

Firstly, we consider the situation where ideal EOS has been used in the definition of the distribution
functions to be implemented in the expressions of $\tau$'s. In this case both a fixed value of coupling, $\alpha_s=0.3$ 
(indicated by the black circles) and the temperature
dependent running coupling $\alpha_s(T)$ (indicated by the orange squares) have been used. The large values of 
$\tau_g$ and $\tau_q$ at lower temperatures indicate the higher values of $\alpha_s(T)$  as compared to fixed, $\alpha_s=0.3$, 
at those range of temperatures. However, at higher temperatures, the much lower values of $\alpha_s(T)$ ($\sim 0.18$ at $T/T_{c}\gtrsim5$),
modulate  the value of thermal relaxation times as compared to the fixed coupling.  Clearly, the logarithmic term,  containing $\alpha_s$,  
playing the key role in determining the behavior of $\tau$'s while  plotted as a function of temperature.
Secondly, in order to visualize the EOS effects in the relaxation times, we introduced the quasiparticle
distribution functions from Eq.(\ref{eq1}) in the expressions of $\tau_g$ and $\tau_q$ along with effective couplings discussed
in Sec. II. Both hard thermal loop (HTL) pQCD EOS for order $O(g^5)$  and $O(g^6\ln(1/g)+\delta)$ 
(EOS1 and EOS2) and the 3-flavor lattice QCD EOS (LEOS) have been considered while implementing the quasiparticle 
properties in the QGP. Finally,  We have observed that at higher temperatures, where the effective couplings ($\alpha_{eff}$) using
the HTL and lattice EOS becomes comparable to  $\alpha_{s}(T)$, the respective plots of $\tau$'s 
almost merge with each other. However, since at $T/T_{c}\sim 8-10$, $\alpha_{eff}$ becomes much smaller ($\sim 0.15$), 
the $\alpha_s^2$ term in the expression of $\tau$'s becomes somewhat predominant to keep the values of $\tau$'s 
a little above the running $\alpha_s(T)$ case. At lower temperatures,  the values remain closer to fixed $\alpha_s$ case as the 
effective coupling becomes closer to $0.3$. Therefore,  it is crucial to mention that throughout the range of temperatures, the logarithmic term containing $\alpha_s$ in
the denominator is playing a predominant role as well as the temperature behavior of the thermal relaxation times of quarks and gluons.

Before presenting the results of electrical conductivity
using the thermal relaxation times including the leading log terms discussed so far, we present $\sigma_{el}$
using the pQCD cross-section taken from Ref.~\cite{Greiner}. The infrared singularity here regularized by the 
Debye mass $m_{D}$ to obtain the cross section as $\frac{d\sigma}{dt}=\frac{d\sigma}{dq_\perp ^2}\simeq
\frac{\alpha_s^2}{(q_{\perp}^2 +m_{D}^2)^2}$, where $q_{\perp}$ is the transverse component of momentum transfer
which for small angle scattering $q_{\perp}^{2}\approx -t$. In a number of recent works,  this cross section
has been used in order to determine the  $\sigma_{el}$ and other transport coefficients as well
~\cite{Greco-1,Greco-2}. We see  that the effect of coupling entering in the expression of differential
scattering cross section as  $\alpha_s^2$  (the predominant logarithmic term $\ln(1/\alpha_s)$ is absent there). 
Employing this cross section we have plotted $\sigma_{el}/T$ as a
function of $T$ in Fig. \ref{sigmapQCD} for a number of EOSs and compare them with some other
estimations of electrical conductivity too. We observe that for ideal EOS
and constant $\alpha_s$, the ratio of electrical conductivity over temperature $\sigma_{el}/T$, is a 
constant over $T$. This is not unexpected as the there is no other temperature dependence 
due to the ideal EOS (the numerical value is close to $\sim 0.06$ as indicated by the
red dasher line). Implementing the running $\alpha_s(T)$ for ideal EOS, we observe that the larger values
of $\alpha_s$ at lower temperatures are 
 decreasing $\sigma_{el}/T$. This is due to the $\alpha_s^2$ term
in the cross section, that is appearing in the denominator of the expression of the $\sigma_{el}$. 
At lower temperatures, the above mentioned  trend agrees with the results of BAMPS (Boltzmann approach for multi-parton
scattering) ~\cite{Xu-Greiner} and Greco {\it et al}. ~\cite{Greco-1}. 
However, the smaller values of $\alpha_s(T)$ at larger temperatures are making $\sigma_{el}/T$ enhanced with respect to
fixed $\alpha_s$ case. Next, we have implemented the effects of EOS in both the distribution
functions and in couplings while determining $\sigma_{el}$. At lower temperatures since $\alpha_{eff}$
is smaller than $\alpha_{s}(T)$, $\sigma_{el}/T$ exhibits larger values demonstrating the equation of state
effects on electrical conductivity. At larger temperatures we can see that these plots almost merges with the one using ideal EOS
and running $\alpha_{s}(T)$, since in those ranges of temperatures $\alpha_{eff}/\alpha_s$ approaches to unity.
Up to $T/T_{c} \sim 2$ we plotted lattice results from Aarts {\it et al}. ~\cite{Arts-2015}.
The quantitative estimations of $\sigma_{el}/T$ with quasiparticle EOSs agrees with the order of magnitude of the 
lattice results.

Next,  we present the results of electrical conductivity using the thermal relaxation times from Eqs. (\ref{taug})
and (\ref{tauq}) including the leading log cross sections in Figs. \ref{sigma-ll-2} and \ref{sigma-ll-3}. 
Due to the predominant contribution from the 
logarithmic term over coupling the magnitude of $\sigma_{el}/T$ becomes quite larger than the pQCD case.
In this case we have plotted $\sigma_{el}/T$ both for 2- and 3-flavors individually for different
EOSs and the 3-flavor case appears to be slightly greater since the quark charge $q_{Q}^2$ in Eq.(\ref{sigma_11})
contains the fractional quark charge of strange quark also. The values of electrical conductivity with quasiparticle
EOS and including $\alpha_{eff}$ as the coupling show smaller values with respect to the one with ideal EOS
and running $\alpha_{s}(T)$ at lower values of temperature due to the leading log effect. However at higher
temperatures the two sets of curves merge with each other due to the fact that at large $T$, $\alpha_{eff}$ approaches
to running $\alpha_{s}(T)$. Although in this case at lower $T/T_{c}$ the lattice data from ~\cite{Arts-2015}
quite underpredicts the results, the  quenched lattice measurement of electrical conductivity from Gupta {\it et al}.
~\cite{Gupta} upto $T/T_{c}\sim 3$ remarkably agrees with the current estimation of $\sigma_{el}$.
For 3-flavor case beyond $T/T_{c}\sim 3$, the estimations of $\sigma_{el}$ is observed to match with the trend
given in Cassing {\it et al}. ~\cite{Cassing} and agrees with their statement that above $T\sim 5T_{c}$
the dimensionless ratio $\sigma_{el}/T$ becomes approximately constant ($\approx0.3$).
In all the above estimations of $\sigma_{el}$ the electronic 
charges are explicitly multiplied using $\frac{e^2}{4\pi}=\frac{1}{137}$.

The diffusion coefficient $D$ estimated from Eq.(\ref{diff-1}) multiplied with $2\pi T$, has been plotted for 
2 and 3-flavor cases as a function of temperature in Figs. \ref{Diff2fig} and \ref{Diff3fig} respectively.
In both the cases the values of diffusion coefficients have been compared with the lattice results
provided by Aarts {\it et al}. ~\cite{Arts-2015}. Although in this case, the $q_{Q}^2$ term containing the flavor
sum over fractional quark charges are absent, the flavor information is embedded in the thermal relaxation times
in $l_{11}$ term. Since for larger quark degeneracy, $\tau_q$ decreases, therefore,  for 3-flavor case the values
of $D$ appear to be smaller than the 2-flavor case. 
Similar to the case of electrical conductivity,
the leading log results for the diffusion coefficient turn out to be much higher 
due to the logarithmic term as compare to situation where only $\alpha_s^2$ is present.
The pQCD results are however closer to the lattice results. The quasiparticle model including the HTL and 
the lattice EOSs is observed to effect the values of $D$ in a significant way. In 3-flavor case,
 we also compare our results with the estimations of $D$ using 
the holographic model from ~\cite{Finazzo}, which are in the range of temperature $0.2-0.4$ GeV
agree well  in the order of magnitude with our pQCD results.
Finally, we have plotted the charge susceptibility as function of  temperature including ideal  EOS and the
EOSs described by EQPM in Fig. \ref{susc}. In lower temperature region, ranging from $0.2-0.35$ GeV, our results show good agreement
with the lattice data from ~\cite{Giudice}. Different EOSs within EQPM  show
discrete effects on $\chi$ . Interestingly, the  ones with LEOS are closer to  the lattice data the most.
 However, like any other quasi-particle model predictions,  the EQPM predictions on the transport coefficients obtained here, show poor matching.
This can perhaps be improved a bit while updating the temperature dependence of effective fugacities in our EQPM with more recent lattice results. Let us enlist the possible route causes of this discrepancy, specifically, in the context of our results on EM transport coefficients   against the 
lattice results  of Aarts {\it et al}. ~\cite{Arts-2015}.
First and foremost reason is the very philosophy to map hot QCD medium effects  in terms of non-interacting/weakly interacting quasi-particle models for the temperatures closer to $T_c$ where the 
interaction measure has a peak  (although, at the level of fitting with the quasi-particle models and yields hot QCD thermodynamics, the matching the reasonably better, however, the very existence of the 
quasi-particle picture is in serious doubt).  Another aspect is clearly the presence of the leading log term in the interaction cross section that follows from the infrared shielding discussed in section II-B. 
We have observed that this term is having the most significant contribution in controlling the temperature behaviour of the relaxation times as well as transport parameters. The resulting enhancement of the
temperature dependence of $\sigma_{el}$ and $D$ over the leading order pQCD results shown, is also responsible for the discrepancies of the current results with the lattice results provided by Aarts {\it et al.}~\cite{Arts-2015}.

There are more recent lattice data with the refined lattices both from HOT QCD collaboration~\cite{hotqcd_eos} and Budapest-Marseille-Wuppertal Collaboration~\cite{borsa_recent}. In fact,  there are 
significant differences on the QCD trace anomaly near , $T_c$ between these two collaborations and also on  $T_c$ itself.  Interpreting them in terms of EQPM  and comparing the 
EM responses studied in this work and other transport coefficients of the QGP would be a matter of future investigations.  Notably, at the level of EQPM, one requires to have lattice results on trace anomaly for the pure glue sector (needed for defining $z_g$).
Once the EQPM description is obtained for above mentioned lattice results, comparison with the predictions from  direct lattice QCD method would prevail  better understanding on the predictive power of the model.  

\section{Conclusion and outlook}
The estimation of the transport coefficients that characterize the response of EM field to the 
electromagnetically charged QGP in the heavy-ion collisions,  with realistic hot QCD/QGP equations 
of state (via their quasi-particle understanding), has led us to very interesting outcomes highlighting the impact of hot QCD medium effects. 
We have investigated the charge transport by determining the 
electrical conductivity of the QGP along with a related phenomenon of charge diffusion in the QGP medium by analyzing the 
charge diffusion coefficient.

The hot QCD medium effects have been  included through the effective quasi-parton distribution functions along with the effective coupling in QCD 
at high temperature. All the transport coefficients that have been investigated in this work, are influenced significantly in the presence of 
hot QCD medium effects coming from the various equations of state under consideration, as compared to the  case 
of ideal equation of state for the QGP.  The results obtained here are 
seen to be consistent with the outcomes of other approaches such as lattice QCD, dynamical quasi-particle models, holographic model based on 
AdS-CFT, transport theory and pQCD based studies discussed in introduction Section. 

The transport coefficients determined in this work  and their temperature dependence could affect the
 quantitative estimates of the signals for the QGP  from heavy ion collisions, particularly, where hydrodynamic simulations are involved.
For example in Refs. ~\cite{Yin,Huot,Ding}, the soft photon emission rate is shown to be linearly dependent upon
electrical conductivity. As a result the hydrodynamic description of the $p_T$ spectra and
elliptic flow of thermal photon and dileptons could be improved by including a realistic temperature dependence of
the electrical conductivity. In that spirit relating the electrical conductivity and charge diffusion coefficients  
to the electromagnetic probes such as dilepton and photon production in 
relativistic heavy-ion collisions, and obtaining the spectra and collective flows would be a matter of immediate future investigations. 

To achieve deeper understanding, the connections 
to the charge fluctuations and directed flow of charged hardrons  in HIC would be another interesting aspect to explore in near future.
In addition, extensions of the present analysis to estimate the 
other transport parameters such as shear and bulk viscosities, thermal conductivity along with generalizations in the  case of anisotropic (momentum) hot QCD medium would 
be another direction where we shall intend to focus on.

\section*{Acknowledgements}
VC would like to acknowledge the funding from Department of Science and Technology, Government of India under 
Inspire Faculty Fellowship, IFA-13/PH-55. SM sincerely acknowledges IIT Gandhanhinagar, India for the Institute postdoctoral fellowship.
People of India are sincerely acknowledged for their generous support for the 
research in basic sciences.

\appendix
\section*{APPENDIX}
\section*{Calculational details of Chapman-Enskog method }
Starting from Eqs.(\ref{trns-1}) and utilizing the thermodynamic identities provided in Eq.(\ref{numberid}),
(\ref{energyid}) and (\ref{velid}) and avoiding the terms containing velocity gradients that gives rise to
viscous phenomena, we land in the following structure of transport equation,

\begin{eqnarray}
&&\frac{1}{T}[p_{k}^{\nu}\sum_{a=1}^{N'-1}(q_{ak}-x_{a}) \{(\nabla_{\mu}\mu_{a})_{P,T}-\frac{h_{a}}{nh}\nabla_{\mu}P\}+\nonumber \\
&&p_{k}^{\nu}\{(p_{k}.u)-h_{k}\} \{\frac{\nabla_{\mu}T}{T}-\frac{\nabla_{\mu}P}{nh}\}+\nonumber \\
&&\{-(p_{k}\cdot u)(p_{k}\cdot E)\frac{\sum_{k=1}^{N}q_{k}n_{k}}{\sum_{k=1}^{N}h_{k}n_{k}}+q_{k}(p_{k}\cdot E)\}]\nonumber \\
&&=-\frac{\omega_{k}}{\tau_{k}}\phi_{k}~.
\label{App-1}
\end{eqnarray}

The first two terms in the left hand side of Eq.(\ref{App-1}) contribute to the diffusion driving force and thermal 
driving forces respectively for a system without the electric field effects. The third term purely arising from the 
effects of the electric field influencing the system. 
Now for generality it is desirable to express the electric filed driven terms in a manner, such that it
resembles the thermal and diffusion driving terms.
In this spirit we can decompose the third term in the following way,

\begin{eqnarray}
 &&\{-(p_{k}\cdot u)(p_{k}\cdot E)\frac{\sum_{k=1}^{N}q_{k}n_{k}}{\sum_{k=1}^{N}h_{k}n_{k}}+q_{k}(p_{k}\cdot E)\}=\nonumber\\
 &&p_{k}^{\nu}\{(p_{k}.u)-h_{k}\}X^{E}_{q\nu}+p_{k}^{\nu}\sum_{a=1}^{N'-1}(q_{ak}-x_{a})X^{E}_{a\nu}~.
\end{eqnarray}
Here we have,
\begin{eqnarray}
 X^{E}_{q\mu}=&&-\frac{1}{h}\sum_{k=1}^{N}x_{k}q_{k}E_{\mu}~,\\
 X^{E}_{k\mu}=&&[q_{k}-q_{N}-\frac{h_{k}-h_{N}}{h}\sum_{l=1}^{N}x_{l}q_{l}]E_{\mu}~,
\end{eqnarray}
as the thermal and diffusion driving forces only due to the influence of the electric field.
 
Applying the decomposition in Eq.(\ref{App-1}) we are finally able to get Eq.(\ref{trns-2}) with
the complete expression of thermal and diffusion driving forces given in Eq.(\ref{forceth}) and (\ref{forcedif})
respectively.


\begin{thebibliography}{99}
\bibitem{expt_rhic}
J. Adams {\it et al.}  (STAR Collaboration), Nucl.  Phys.  A {\bf 757}, 102 (2005);
 K. Adcox {\it et al.} PHENIX Collaboration, Nucl. Phys.  A {\bf 757}, 184 (2005);
 B.B. Back {\it et al.} PHOBOS Collaboration, Nucl. Phys. A  {\bf 757}, 28 (2005);
 I. Arsene {\it et al.} BRAHMS Collaboration, Nucl. Phys. A {\bf 757}, 1 (2005). 

\bibitem{expt_lhc}
K. Aamodt {\it et al.} (The Alice Collaboration), Phys. Rev.
Lett. {\bf 105}, 252302 (2010);
 Phys. Rev.  Lett. {\bf 105}, 252301 (2010); Phys. Rev. Lett. {\bf 106}, 032301
(2011).

\bibitem{kss}
  P. Kovtun, D. T. Son, and A. O. Starinets, Phys. Rev. Lett.
{\bf 94}, 111601 (2005). 
 
  
\bibitem{lowvisc} L. P. Csernai, J. I. Kapusta, and L. D. McLerran, Phys. Rev.
Lett.  {\bf 97}, 152303 (2006).
  
  \bibitem{lacey}
Roy A. Lacey {\it et al.},   Phys. Rev.  Lett. {\bf 98}   092301 (2007).

\bibitem{bulk} D. Kharzeev and K. Tuchin, J. High Energy Phys.  {\bf 09}
(2008) 093; F. Karsch, D. Kharzeev, and K. Tuchin, Phys. Lett.  {\bf B 663}, 217 (2008); P. Romatschke and D. T. Son, Phys. Rev. D {\bf 80}, 065021 (2009); G. D. Moore and O. Saremi, J. High Energy Phys. {\bf 09} (2008) 015; C. Sasaki and K. Redlich, Phys. Rev. C {\bf 79}, 055207 (2009); Nucl. Phys. A {\bf 832}, 62 (2010). 
  
   \bibitem{em_rhic}
   B.G. Zakharov, 
   Phys. Lett. {\bf B 737} (2014) 262;  K. Tuchin, 
     Phys. Rev. C 91, 064902 (2015).
     
  \bibitem{em_decay}  K. Tuchin, 
  Adv. High Energy  Phys.  {\bf 2013} (2013) 490495.
  
  \bibitem{lerry}
  L.  McLerran and V. Skokov, 
  Nucl. Phys.  A {\bf 929}  (2014) 184.
  
  \bibitem{kharz} U. G\"{u}rsoy, D. Kharzeev and K. Rajagopal,
  Phys. Rev. C {\bf 89} (2014) 054905;
 \bibitem{satw} D. Satow, 
  Phys. Rev. D  {\bf 90 } (2014) 034018.  
 
\bibitem{charge_fluc}
 B. Ling, T. Springer and M. Stephanov, 
   Phys.  Rev.  C  {\bf 89} (2014) 064901.
   
   \bibitem{flow} 
  Y. Hirono, M. Hongo and T. Hirano, 
   Phys. Rev. C {\bf 90} (2014) 021903. 
   
   \bibitem{Greif_heat}
   M.~Greif, F.~Reining, I.~Bouras, G.~S.~Denicol, Z.~Xu and C.~Greiner,
  Phys.\ Rev.\ E {\bf 87} (2013) 033019
  doi:10.1103/PhysRevE.87.033019
  [arXiv:1301.1190 [hep-ph]].
  
  \bibitem{Dobado}
   A.~Dobado, F.~J.~Llanes-Estrada and J.~M.~Torres Rincon,
  hep-ph/0702130 [HEP-PH].
  
  \bibitem{green-kubo}
  M. S. Green, 
  J. Chem. Phys. {\bf 22}, 398 (1954);
   R. Kubo, 
  J. Phys. Soc. Japan {\bf 12} (1957) 570.
  
  \bibitem{Arts-2007}
  G.~Aarts, C.~Allton, J.~Foley, S.~Hands and S.~Kim,
  Phys.\ Rev.\ Lett.\  {\bf 99} (2007) 022002.
  
  \bibitem{Amato}
  A.~Amato, G.~Aarts, C.~Allton, P.~Giudice, S.~Hands and J.~I.~Skullerud,
  Phys.\ Rev.\ Lett.\  {\bf 111} (2013) no.17,  172001.
  
  \bibitem{Arts-2015}
   G.~Aarts, C.~Allton, A.~Amato, P.~Giudice, S.~Hands and J.~I.~Skullerud,
  JHEP {\bf 1502} (2015) 186
  doi:10.1007/JHEP02(2015)186
  [arXiv:1412.6411 [hep-lat]].
  
  \bibitem{Gupta}
   S.~Gupta,
  Phys.\ Lett.\ B {\bf 597} (2004) 57
  doi:10.1016/j.physletb.2004.05.079
  [hep-lat/0301006].
  
  \bibitem{Giudice}
  P.~Giudice, G.~Aarts, C.~Allton, A.~Amato, S.~Hands and J.~I.~Skullerud,
  PoS LATTICE {\bf 2013} (2014) 492
  [arXiv:1309.6253 [hep-lat]].
  
  \bibitem{Brandt}
  B.~B.~Brandt, A.~Francis, H.~B.~Meyer and H.~Wittig,
  PoS ConfinementX {\bf } (2012) 186
  [arXiv:1302.0675 [hep-lat]].
  
  \bibitem{Francis}
  A.~Francis and O.~Kaczmarek,
  Prog.\ Part.\ Nucl.\ Phys.\  {\bf 67} (2012) 212
  doi:10.1016/j.ppnp.2011.12.020
  [arXiv:1112.4802 [hep-lat]].
  
  \bibitem{Buividovich}
  P.~V.~Buividovich, M.~N.~Chernodub, D.~E.~Kharzeev, T.~Kalaydzhyan, E.~V.~Luschevskaya and M.~I.~Polikarpov,
  Phys.\ Rev.\ Lett.\  {\bf 105} (2010) 132001
  doi:10.1103/PhysRevLett.105.132001
  [arXiv:1003.2180 [hep-lat]].

  \bibitem{Xu-Greiner}
 M. Greif, I. Bouras, and C. Greiner,
 {Phys. \ Rev. \ D {\bf 90}, 094014 (2014)}~.

  \bibitem{Greco-1}
   A.~Puglisi, S.~Plumari and V.~Greco,
  Phys.\ Lett.\ B {\bf 751} (2015) 326.
  
  \bibitem{Greco-2}
  A.~Puglisi, S.~Plumari and V.~Greco,
  Journal of Physics: Conference Series 612 (2015) 012057.
  
  \bibitem{Cassing}
  W.~Cassing, O.~Linnyk, T.~Steinert and V.~Ozvenchuk,
  Phys.\ Rev.\ Lett.\  {\bf 110} (2013) no.18,  182301
  doi:10.1103/PhysRevLett.110.182301
  [arXiv:1302.0906 [hep-ph]].
  
  \bibitem{Cassing1}
   T.~Steinert and W.~Cassing,
  Phys.\ Rev.\ C {\bf 89} (2014) no.3,  035203
  doi:10.1103/PhysRevC.89.035203
  [arXiv:1312.3189 [hep-ph]].
  
  \bibitem{Qin} 
  S.~X.~Qin,
  Phys.\ Lett.\ B {\bf 742} (2015) 358
  doi:10.1016/j.physletb.2015.02.009
  [arXiv:1307.4587].
  
  \bibitem{Finazzo}
  S.~I.~Finazzo and R.~Rougemont,
  Phys.\ Rev.\ D {\bf 93} (2016) no.3,  034017
  doi:10.1103/PhysRevD.93.034017
  [arXiv:1510.03321 [hep-ph]].
  
  \bibitem{Binoy}
  P.~K.~Srivastava, L.~Thakur and B.~K.~Patra,
  Phys.\ Rev.\ C {\bf 91} (2015) no.4,  044903
  doi:10.1103/PhysRevC.91.044903
  [arXiv:1501.03576 [hep-ph]].
  
  \bibitem{Fraile}
  D.~Fernandez-Fraile and A.~Gomez Nicola,
  Phys.\ Rev.\ D {\bf 73} (2006) 045025
  doi:10.1103/PhysRevD.73.045025
  [hep-ph/0512283].
  
  \bibitem{Denicol}
    M.~Greif, C.~Greiner and G.~S.~Denicol,
  Phys.\ Rev.\ D {\bf 93} (2016) no.9,  096012
  doi:10.1103/PhysRevD.93.096012
  [arXiv:1602.05085 [nucl-th]].
  
  \bibitem{chandra_quasi1}
  Vinod Chandra, R. Kumar, V. Ravishankar, Phys. Rev.  C {\bf 76}  (2007) 054909, [Erratum: Phys. Rev.  C {\bf 76}, 069904  (2007).
  Vinod Chandra, A. Ranjan, V. Ravishankar,  Eur. Phys. J.  A {\bf 40}, 109-117  (2009).
    
  \bibitem{chandra_quasi2} Vinod Chandra, V. Ravishankar  
  Phys. Rev.  D {\bf 84}, 074013   (2011). 
  
  \bibitem{Bluhm}
  M.~Bluhm, B.~Kampfer and K.~Redlich,
  Phys.\ Rev.\ C {\bf 84} (2011) 025201
 
 \bibitem{chandra_eta} 
 Vinod Chandra, V. Ravishankar, Eur.  Phys.  J. C  {\bf 64}, 63-72  (2009) ; {\it ibid.} C {\bf 59}, 705-714  (2009).
  
\bibitem{chandra_etazeta}
Vinod Chandra, Phys. Rev. D {\bf 86} 114008 (2012); {\it ibid.}, D {\bf 84}, 094025  (2011).
 

\bibitem{PJI}
P. Chakraborty, J. I.  Kapusta , 
Phys.  Rev.  C  {\bf 83},  014906 (2011).

 \bibitem{Mkap} 
 Albright and J. I. Kapusta,  Phys. Rev. C {\bf 93}, 014903 (2016).

\bibitem{Greco} A. Puglisi, , S. Plumari,  V. Greco, Phys. Lett.  B {\bf 751}, 326-330 (2015).


\bibitem{Ryu}
S.  Ryu, J. -F.  Paquet, C.  Shen, G. S. Denicol, B. Schenke, S. Jeon,  C. Gale, Phys.  Rev.   Lett.  {\bf 115}, 132301 (2015).

\bibitem{Denicol1}
G.  Denicol, A.  Monnai, B.  Schenke,   Phys. Rev.  Lett. {\bf 116}, 212301 (2016).


  \bibitem{effmass1}
 A. Peshier  {\it et. al}, Phys.  Lett.  {\bf B 337}, 235 (1994); Phys.
Rev. D {\bf 54}, 2399 (1996).
 
 \bibitem{effmass2} A. Peshier, B. K\"{a}mpfer, G. Soff, Phys. Rev. C {\bf 61},
045203 (2000); Phys. Rev.  D {\bf 66}, 094003 (2002); V. M. Bannur, Phys. Rev. C {\bf 75}, 044905 (2007); {\it ibid}. C {\bf 78}, 045206 (2008); JHEP {\bf 0709}, 046 (2007); A. Rebhan, P. Romatschke, Phys. Rev. D {\bf 68}, 0250022 (2003); 
 M. A. Thaler, R. A. Scheider, W. Weise, Phys. Rev. C {\bf 69}, 035210 (2004);  K. K. Szabo, A. I. Toth, JHEP  {\bf 0306}, 008 (2003). 
 
 \bibitem{polya}
 M. D\'Elia, A. Di Giacomo, E. Meggiolaro, Phys. Lett.  {\bf B 408}, 315 (1997); Phys. Rev. D {\bf  67}, 114504 (2003); P. Castorina, M. Mannarelli, Phys. Rev. C {\bf 75}, 054901 (2007);  Phys.  Lett.  {\bf B 664}, 336 (2007)., 
 Paolo Alba {\it et al.}, Nucl. Phys.  A {\bf 934}, 41-51 (2014).  
  
  \bibitem{cheng}
  M. Cheng {\bf et. al}, Phys. Rev.  D {\bf 77} ,  014511  (2008). 
  
  \bibitem{leos1_lat}
M. Cheng {\it et al.}, Phys. Rev. D  {\bf 77}, 014511 (2008); 
 A. Bazavov {\it et al.}, Phys. Rev. D {\bf 80}, 014504 (2009); 
M. Cheng {\it et al.}, Phys. Rev. D {\bf 81}, 054504 (2010); 
S. Borsanyi {\it et al.} , J. High Energy Phys. {\bf 11} (2010) 077; Y.
Aoki {\it et al.}, J. High Energy Phys. {\bf 01} (2006) 089; J. High
Energy Phys. {\bf 06}  (2009) 088; 
S. Borsanyi {\it et al.}, J. High Energy Phys. {\bf 09}  (2010) 073.

 \bibitem{zhai} 
P.  Arnold and Chengxing Zhai, Phys.  Rev.  D {\bf  50},  7603;(1994);
{\it ibid} {\bf 51},   1906( 1995).

\bibitem{kastening}
Chengxing Zhai and B. Kastening, Phys. Rev. D {\bf 52}, 7232. (1995).

\bibitem{kaj}
K. Kajantie, M. Laine, K. Rummukainen and Y. Schroder,
Phys. Rev. D {\bf 67},  105008 (2003).  

 \bibitem{qcd_coupling}  M. Laine and Y. Sch\"{o}der,  J HEP {\bf 0503} (2005) 067 [{\tt hep-ph/0503061}].  
 
   
\bibitem{pnjl}
A. Dumitru, R. D. Pisarski, Phys. Lett.  {\bf B 525}, 95 (2002); K. Fukushima, Phys.  Lett.  {\bf B 591}, 277 (2004); S. K. Ghosh {\it et. al}, Phys.  Rev.  D {\bf 73}, 114007 (2006); H. Abuki, K. Fukushima, Phys. Lett.  {\bf B 676}, 57 (2006); 
H. M. Tsai, B. M\"{u}ller,  J. Phys.  G  {\bf 36}, 075101 (2009).

\bibitem{dmass1}
Kelly  {\it et. al}, Phys. Rev. Lett.  {\bf 72}, 3461 (1994), Phys. Rev. D {\bf 50}, 4209 (1995).
\bibitem{dm_rev1}
 Daniel F. Litim, C. Manual, Phys. Rep.  {\bf 364}, 451 (2002).
\bibitem{dm_rev2} 
  J. P. Blaizot, E. Iancu, Phys. Rep. {\bf 359}, 355 (2002).
  
\bibitem{chandra_dilep}
Vinod Chandra and  V. Sreekanth,  Phys.Rev. D {\bf 92}, 094027 (2015); Vinod Chandra and V. Sreekanth,  {\tt arXiv:1602.07142 [nucl-th]}. 

\bibitem{chandra_hq1}
 Vinod Chandra, V. Ravishankar,  Nucl. Phys.  A {\bf 848}, 330-340  (2010).

\bibitem{chandra_hq2}
  Vinod Chandra, S. K. Das, Phys. Rev.  D {\bf 93},  094036   (2016);  S. K. Das, Vinod Chandra, Jan-e-Alam, J. Phys. G {\bf 41}, 015102   (2013).
 
\bibitem{AMY2}
P.~B.~Arnold, G.~D.~Moore and L.~G.~Yaffe,
  JHEP {\bf 0011} (2000) 001
  doi:10.1088/1126-6708/2000/11/001
  [hep-ph/0010177].
  
\bibitem{Zhang}

  X.~F.~Zhang and W.~Q.~Chao,
  Nucl.\ Phys.\ A {\bf 628} (1998) 161~.
 
 \bibitem{Combridge}
  B.~L.~Combridge, J.~Kripfganz and J.~Ranft,
  Phys.\ Lett.\ B {\bf 70} (1977) 234.
  
  \bibitem{Thoma}
  M.~H.~Thoma,
  Phys.\ Rev.\ D {\bf 49} (1994) 451~.
  
  \bibitem{Hosoya}
  A.~Hosoya and K.~Kajantie,
  Nucl.\ Phys.\ B {\bf 250} (1985) 666.
  
\bibitem{Degroot}
S.~R.~De Groot, W.~A.~Van Leeuwen and C.~G.~Van Weert,
{\it Relativistic Kinetic Theory, Principles And Applications}
{\it  Amsterdam, Netherlands: North-holland (1980)}. 


\bibitem{Greif}
M. Greif,  {\it Electric Conductivity of the Quark-Gluon Plasma},
{\it Master Thesis, Goethe Universit\"{a}t, Frankfurt Am Main (25 May, 2014)}.


\bibitem{Baym}
G.~Baym, H.~Monien, C.~J.~Pethick and D.~G.~Ravenhall,
  Nucl.\ Phys.\ A {\bf 525} (1991) 415C.
  
 \bibitem{Greiner}
 Z.~Xu and C.~Greiner,
  Phys.\ Rev.\ C {\bf 71} (2005) 064901
  doi:10.1103/PhysRevC.71.064901
  [hep-ph/0406278]. 
  
 \bibitem{hotqcd_eos}
A.  Bazavov {\it et al.},   Phys.  Rev.  D {\bf 90}, 094503 (2014).


\bibitem{borsa_recent}
S.  Borsanyi, Z.  Fodor, C. Hoelbling, S. D. Katz, S. Krieg, K. K. Szabo,   Phys.  Lett.  {\bf B 370}, 99-104,  (2014).


  \bibitem{Yin}
  Y.~Yin,
  Phys.\ Rev.\ C {\bf 90} (2014) no.4,  044903.
  
  \bibitem{Huot}
  S.~Caron-Huot, P.~Kovtun, G.~D.~Moore, A.~Starinets and L.~G.~Yaffe,
  JHEP {\bf 0612} (2006) 015.
  
  \bibitem{Ding}
   H.-T.~Ding, A.~Francis, O.~Kaczmarek, F.~Karsch, E.~Laermann and W.~Soeldner,
  Phys.\ Rev.\ D {\bf 83} (2011) 034504.
  
\end{thebibliography}
\end{document}